# Predicting and Optimizing for Energy Efficient ACMV Systems: Computational Intelligence Approaches

*Deqing Zhai* [a, *], *Yeng Chai Soh* [a]

[a]*School of Electrical and Electronic Engineering, 50 Nanyang Ave, Nanyang Technological University, Singapore 639798*

## Abstract

In this study, a novel application of neural networks that predict thermal comfort states of occupants is proposed with accuracy over 95%, and two optimization algorithms are proposed and evaluated under two real cases (general offices and lecture theatres/conference rooms scenarios) in Singapore. The two optimization algorithms are Bayesian Gaussian process optimization (BGPO) and augmented firefly algorithm (AFA). Based on our earlier studies, the models of energy consumption were developed and well-trained through neural networks. This study focuses on using novel active approaches to evaluate thermal comfort of occupants and so as to solves a multiple-objective problem that aims to balance energy-efficiency of centralized air-conditioning systems and thermal comfort of occupants. The study results show that both BGPO and AFA are feasible to resolve this no prior knowledge-based optimization problem effectively. However, the optimal solutions of AFA are more consistent than those of BGPO at given sample sizes. The best energy saving rates (ESR) of BGPO and AFA are around -21% and -10% respectively at energy-efficient user preference for both Case 1 and Case 2. As a result, an potential benefit of S$1219.1 can be achieved annually for this experimental laboratory level in Singapore.

* Corresponding author
Email address: dzhai001@e.ntu.edu.sg

# Introduction

According to the Paris Protocol in the Paris Climate Conference 2015, the targets of global climate change were settled. A target of keeping global temperature increment less than 2℃ by year 2020 was agreed globally. Singapore government also actively took part in this agreement and made a pledge with a target of “7% - 11%” global warming emission reduction under the conditions “business-as-usual” by the year of 2020 [1].

One of the major energy consuming factors is the ever increasing number of buildings nowadays. According to the statistics of United States and European Union, buildings consume about 40% of the total energy delivered from power plants [2, 3]. About 40%-60% of the energy consumed by buildings is attributed to Heating, Ventilation and Air-Conditioning (HVAC) or Air-Conditioning and Mechanical Ventilation (ACMV) systems [3, 4]. In addition, the HVAC systems directly control and maintain indoor environmental conditions, which can significantly impact indoor thermal comfort sensations of occupants. The indoor thermal comfort sensations can also influence the work productivity and health of occupants significantly [5]. Moreover, people generally spend about 90% of daily time staying indoors according to a study by the Berkeley National Laboratory [6]. Thus, the obvious importance to maintain healthy and comfortable indoor conditions is clearly drawn. Due to heavy energy consumption of HVAC systems, it is important to optimally balance the HVAC systems' energy-efficiency and occupants' indoor thermal comfort for the sake of energy saving, work productivity and health.

Indoor thermal comfort is a very subjective term. Thermal sensations are subject to variation with respect to different thermal conditions and occupants. Thus the challenging and importance of studying in this field are well established. There are two basic categories among all the methodologies according to literatures. Passive approaches are systematic ways that can be dated back to 1970s, and named as Predicted Mean Vote (PMV) model [7]. P.O. Fanger developed this model through a series of physical laws of heat transfer based on different objective parameters. Therefore, this subjective term was translated into a quantified term via these objective parameters. From then on, there were many successors to further push the research boundaries forward [8]. Besides using physical laws, it has been proved that active approaches directly using large amount of occupant physiological data can also predict thermal comfort of occupants effectively [9, 10]. Nowadays, the most adaptable and significant physiological parameters of predicting thermal comfort sensations are skin temperature [11, 12].

Since this study is an advancement of our previous works [13, 14], the machine learning algorithms and augmented firefly algorithm (AFA) have been thoroughly studied. Hence this study focuses on 1) active approaches on modeling thermal comfort of occupants through machine learning, 2) evaluations and comparisons between two optimization algorithms in balancing energy-efficiency and thermal comfort under two cases with different user-preference weight coefficient from $\lambda \in [0, 1]$. In addition, the thermal comfort of occupants is using active approaches proposed in this study as well. The experimental testbed has been established in the laboratory named \Air Conditioning and Mechanical Ventilation (ACMV) systems" in Nanyang Technological University, Singapore. The reason for selecting these two particular optimization algorithms is that these two algorithms can both search for global optima through limited and online updating datasets that aim to resolve this nature of optimization problem.

The rest of this paper is organized as follows: Section 2 briefly reviews the existing literature on related works. Section 3 describes the methodologies of this study. Section 4 presents, analyzes and discusses the experimental results. Section 5 draws the conclusive remarks and highlights some limitations and future outlooks.

**Table 1: Nomenclature**

| Symbol | Description | Remarks |
|---|---|---|
| $\omega$ | A Column Vector of $\omega^{(1)}$, $\omega^{(2)}$ and $\omega^{(3)}$ | — |
| $\omega^{(1)}$ | Supply-Air Fan Operating Frequency | $Hz$ |
| $\omega^{(2)}$ | Compressor Operating Frequency | $Hz$ |
| $\omega^{(3)}$ | Water Pump Operating Frequency | $Hz$ |
| $T_s$ | Skin Temperature | $°C$ |
| $T_s'$ | Gradient of Skin Temperature | $°C/min$ |
| $T_{s_norm}$ | Normalized Skin Temperature | $°C$ |
| $T_{s_norm}'$ | Normalized Gradient of Skin Temperature | $°C/min$ |
| $T_{sampling}$ | Wearable Device Sampling Period | $min$ |
| $\phi$ | Relative Humidity | $\%$ |
| $M$ | Metabolic Rate | $W/m^2$ |
| $I_{cl}$ | Clothing Insulation Factor | $Km^2/W$ |
| $\lambda$ | User-Preference Weight Coefficient | — |
| $E$ | Energy Consumption of ACMV Systems | $kWh$ |
| $PTS$ | Predictive Thermal State of Occupants | $(-1, 0, +1)$ |
| $\alpha$ | Coefficient of Distance | — |
| $\beta$ | Coefficient of Randomness | — |
| $\gamma$ | Coefficient of Vortex | — |
| $\epsilon$ | Randomness Distribution | Gaussian (0,1) |
| $\boldsymbol{\theta}^{(1)}$ | Input - Hidden-Layer weight parameters | — |
| $\boldsymbol{\theta}^{(2)}$ | Hidden-Layer - Output weight parameters | — |

## Literature Review

This section reviews and discusses the state-of-the-art methodologies, backgrounds and findings relevant to our work. Michailidis et al. [15] developed L4GPCAO optimizing controller which has properties of model-free, plug-n-play and dynamic. The objective function is also based two aspects that are energy consumption and thermal comfort with an introduced weight parameter. However, the weight parameter in this objective function is fixed once optimizer defined. Similarly, Wu et al. [16] also developed Net-Zero Energy Buildings (NZEB) models with different ventilation, dehumidification and heat pumps options. The different ventilations can result in building's energy reduction by 7.5% and 9.7% complying with thermal comfort of occupants under Fanger's PMV thermal comfort model. The different dehumidification options lead to building's energy reduction by 3.9%, and the different heat pump options save about 13.1% and 14.7% of building's energy for 2 boreholes and 3 boreholes respectively. In the study of He et al., higher heating or cooling capacity of HVAC systems do not absolutely provide better indoor comfort sensations [17]. Combined with the study of Fathollahzadeh et al., the Predicted Mean Vote (PMV) model is validated as an exemplified approach to predict indoor thermal comfort and psychological acceptance of occupants [18]. Besides the passive approaches of PMV models on evaluating thermal comfort states of occupants, recent researches also turn the direction on skin temperatures [12]. The active approaches directly translate occupant skin information into thermal comfort sensations by numerical expressions or machine learning techniques [11, 10, 19]. In the studies of Wang et al. and Cetin et al., the experimental results validated that ambient air temperature plays the most important role in indoor thermal comfort sensations of occupants [20, 21]. Moreover, the measurement parameters in our studies are primarily ambient air temperature, ambient air velocity and ambient air relative humidity. In the study of Wong et al., a Bayesian approach of predicting thermal comfort of occupants was proposed. The maximum likelihood estimation was close to actual percentage dissatisfied (APD) in large sample sizes, and the Bayesian estimation was close to Fanger's prediction in small sample sizes [22]. Therefore, we extend the applications of Bayesian approaches in this study as Bayesian optimization [23]. In the studies of Ascione et al. and Dussault et al., a Model-based Predictive Control (MPC) with multiple objectives Genetic Algorithm (GA) optimizations had been developed for HVAC systems [24, 25]. In the studies of Chen et al. and Pritoni et al., the experimental results revealed that occupants' feedbacks in control ACMV systems could provide an energy-efficient way of operating the systems with considerations of maintaining indoor thermal comfort of occupants [26, 27]. Therefore, a feedback loop has been proposed in our work for optimizing and controlling the ACMV systems as presented in Figure 1. In the study of Kim et al., weight coefficients of multiple objectives optimization were introduced [28]. Guided by Kim et al.'s study, the introduction of user-preference weight coefficients is adopted in our work for leveraging energy consumption of HVAC systems and thermal comfort of occupants. Following the methodology directions of these previous works, our present study focuses on modeling developments through neural networks on energy consumption and thermal comfort sensations, and optimizing objective functions through two algorithms namely Bayesian Gaussian Process Optimization (BGPO) and Augmented Firefly Algorithm Optimization (AFA).

## Methodology

The heating, ventilation and air-conditioning (HVAC) systems in this study is customized for Singapore tropical climate. Therefore, only cooling capacity is under evaluations and the testbed laboratory are named Air-Conditioning and Mechanical Ventilation (ACMV) systems, and the whole systems are illustrated in Figure 2. The experimental room is called thermal laboratory, which is isolated from open outside environment. Therefore, the indoor environment of this experimental room can be uniformly and directly conditioned by the ACMV systems. There are three parts in the ACMV systems, namely Air Handling Unit (AHU), Water Chiller Unit (WCU) and Liquid Dehumidification Unit (LDU). In this study, the LDU is not taken into consideration due to our limited study scope. Therefore, the main energy consuming components are supply-air fan motor, compressor, water pump and condenser in this study.

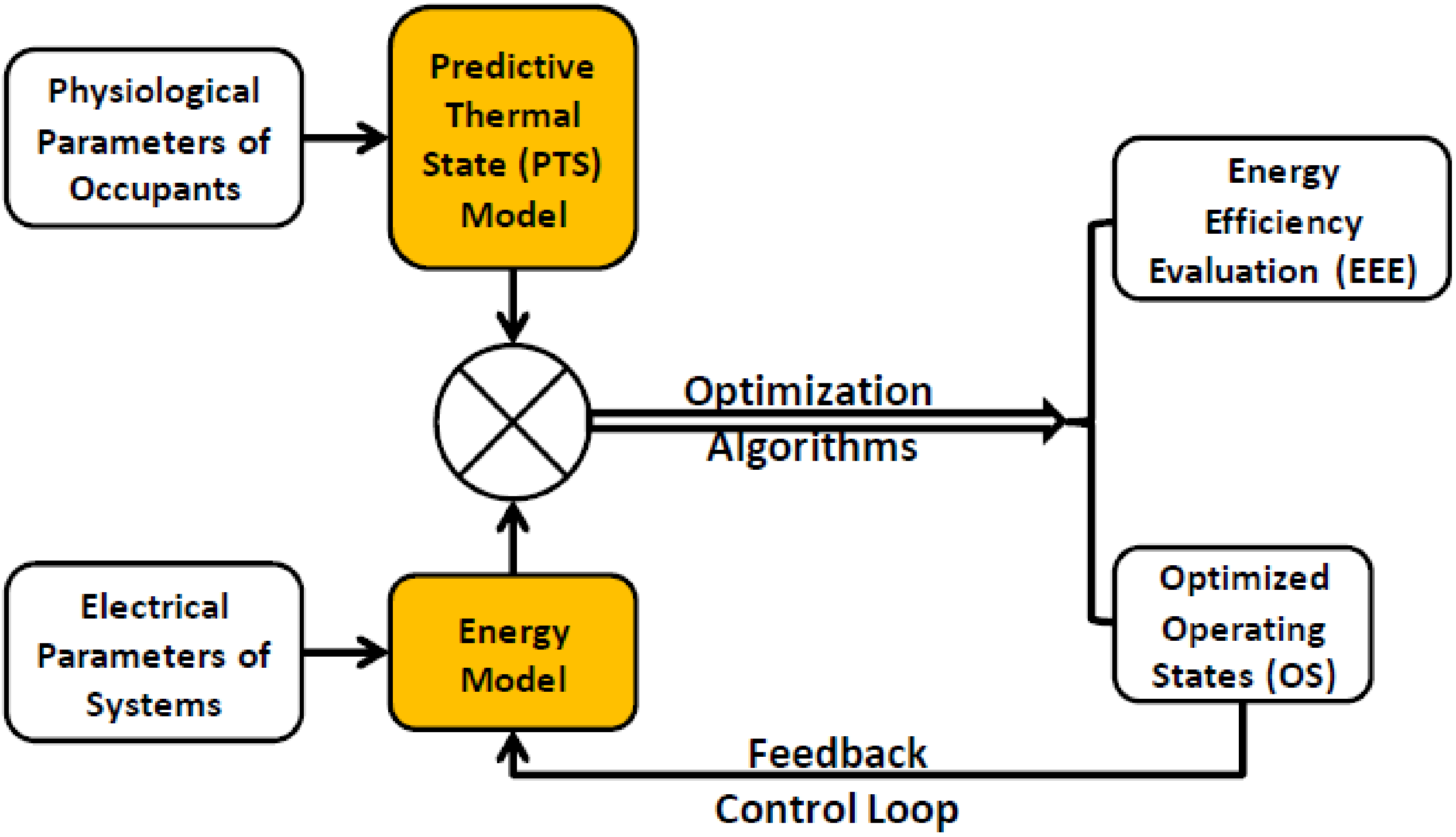

Figure 1: Overall Analytic Diagram of Study

The neural network models of energy consumption, ambient air temperature and air velocity have been developed and well trained based on our earlier studies [13, 14, 29, 30], and the model evaluations are tabulated in Table 5-7. The overall analytic model is shown in Figure 1. Given the knowledge of occupant and environmental parameters, active approaches had been applied for evaluating indoor thermal comfort status of a single particular occupant. An objective function is linearly formulated with a user-preference weight coefficient (λ) associated with both energy consumption level and indoor thermal comfort sensation. Given the formulated objective function, two optimization algorithms are ready for energy efficiency evaluations (EEE). Both optimization algorithms perform a search for global optima given the boundary constraints and limited sample sizes. The evaluation results with different user-preference weight coefficients are tabulated and illustrated in the following sections in detail.

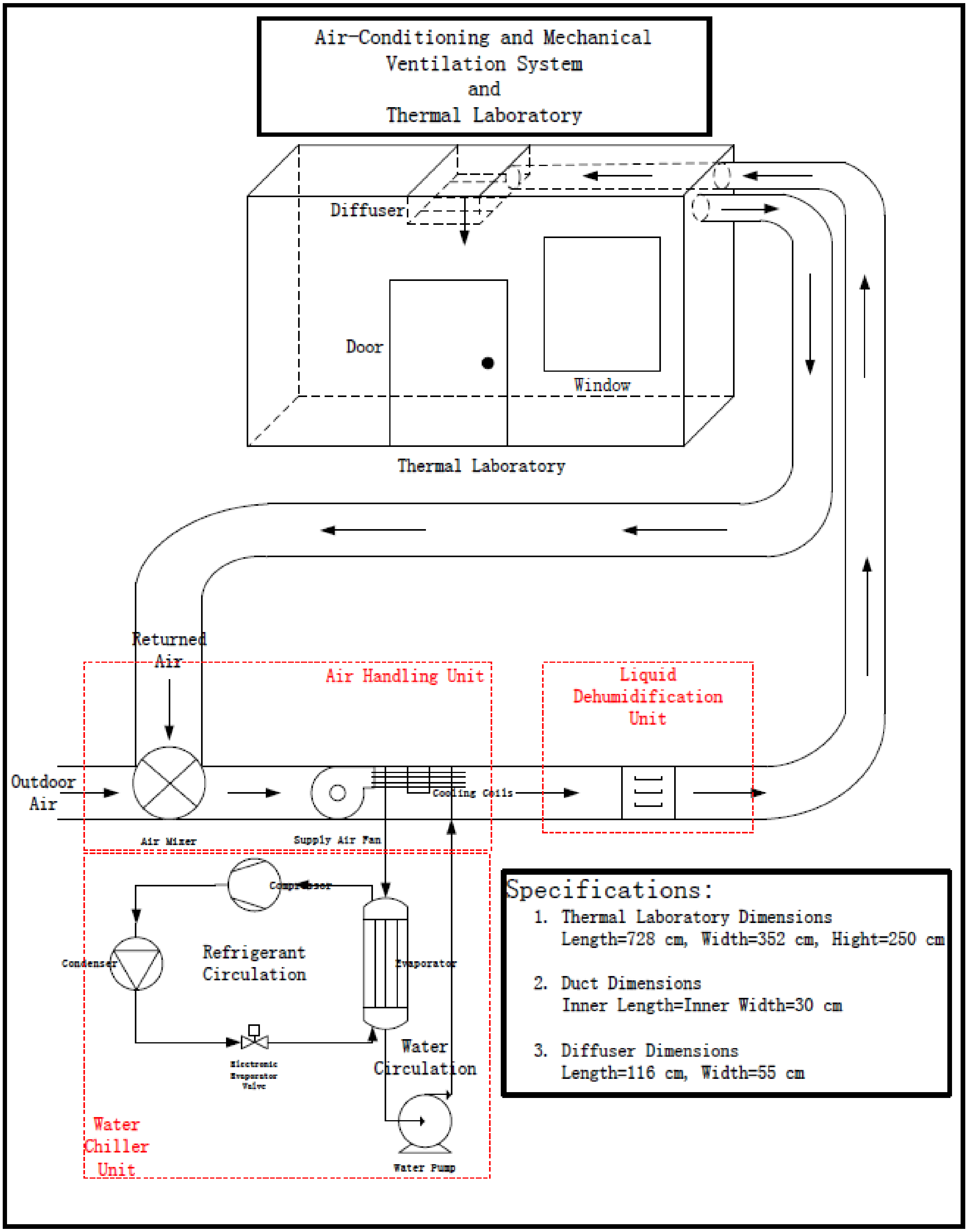

Figure 2: Air-Conditioning and Mechanical Ventilation (ACMV) Systems

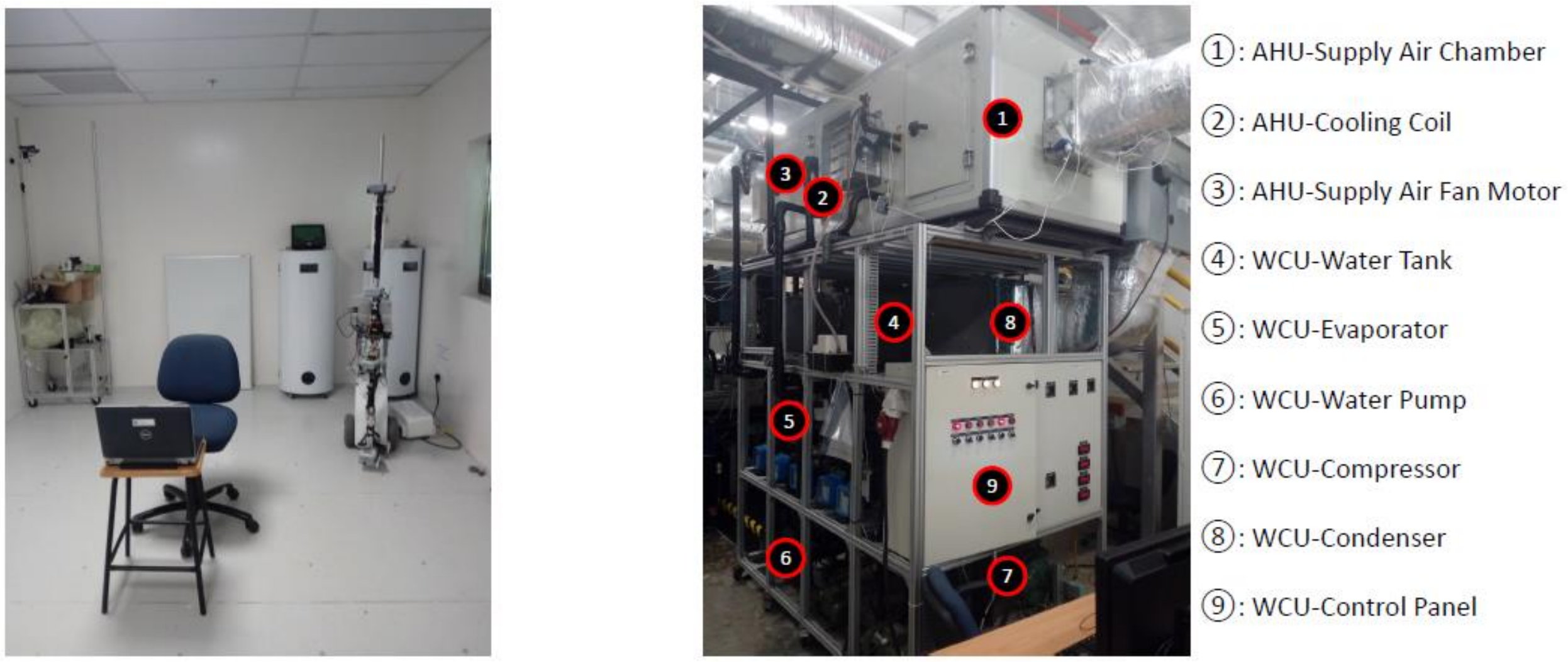

(a) Thermal Laboratory (b) AHU and WCU

**Figure 3: Experimental Environment and Equipment**

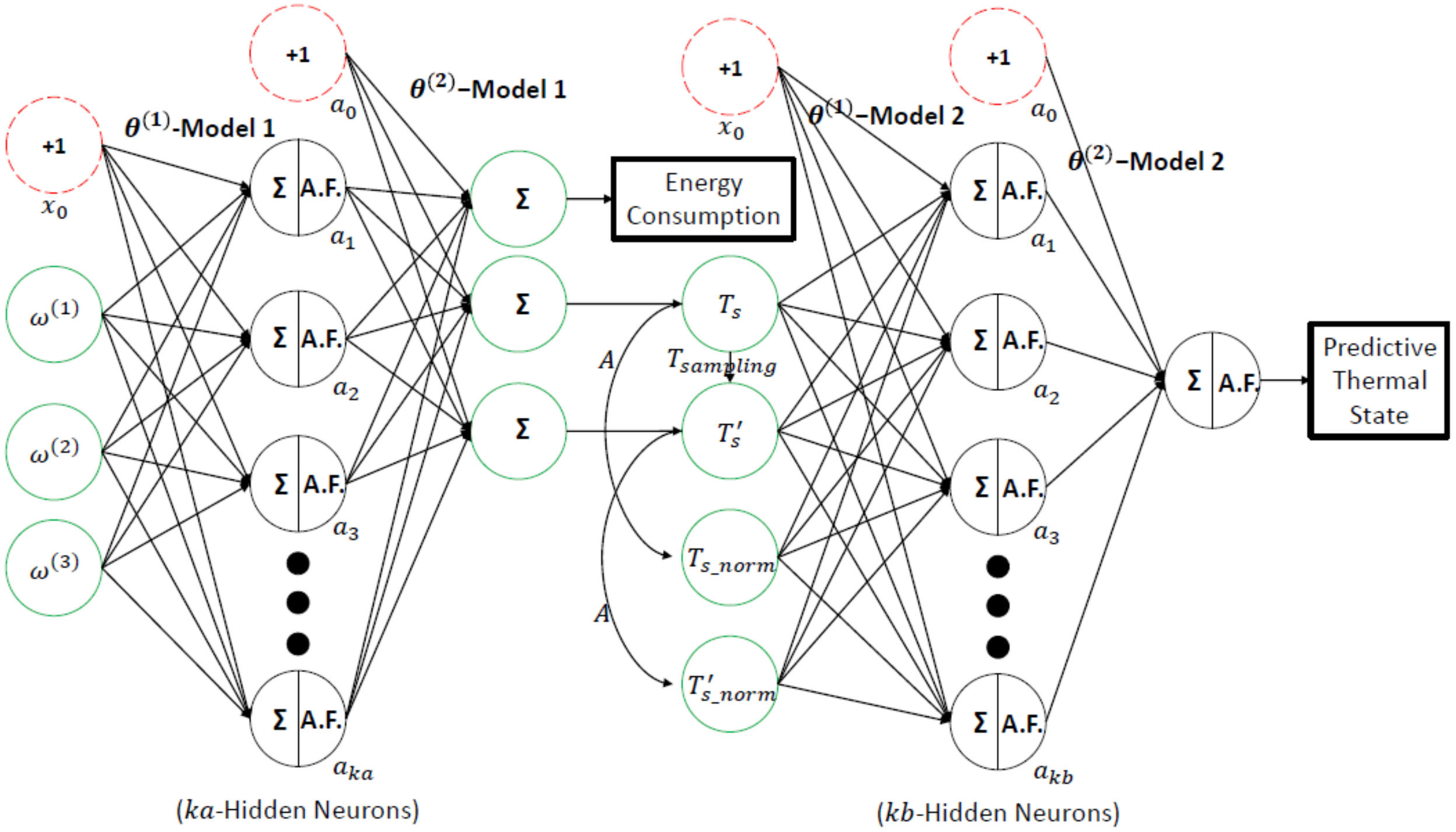

**Figure 4: Neural Networks Modeling Topology**

## Modeling of Energy Consumption

Based on our earlier studies, the models of energy consumption, ambient air temperature and air velocity have been developed and refined through neural networks with respect to operating frequencies of Air-Conditioning and Mechanical Ventilation (ACMV) systems [13]. The model topology is illustrated in Figure 4 and the model evaluations are presented. Sigmoid function is selected as the activation function for each neuron of the neural networks. It is to be noted that the raw data of both training and testing datasets were subjected to an introduced initial normalization process to translate into standard datasets, and the raw data inputs are normalized through standard Gaussian.

## Modeling of Thermal Comfort

In 1970s, P.O. Fanger had developed a predicted mean vote (PMV) model, which is a systematical way to evaluate occupant thermal comfort based on two types of parameters, namely occupant and environmental parameters [7]. The occupant parameters are metabolic rates (or activity levels) and clothing insulation factors of an occupant. The environmental parameters are air temperature, air velocity, air relative humidity and mean radiant temperature in the occupant's proximity. Due to the complexity of measurements in Fanger's model, active approaches are validated and applied in this study with skin temperature information of occupants. The skin temperature is selected only on backside of hands according to relevant researches and studies. There are two case studies under discussions, which are general office and lecture theatre/conference room. The measurement instrumentations are listed in Table 2.

Table 2: Measurement Instrumentations

| Parameter | Model | Range | Accuracy |
|---|---|---|---|
| Air Velocity | TSI Air Velocity Transducer 8475 | 0.05 - 2.5 $m/s$ | $\pm$ 3.0% |
| Air Temperature | EE21-FT3B56/T07 | 0 - 60 $°C$ | $\pm$ 2.0% |
| Air Relative Humidity | EE21-FT3B56/T07 | 0 - 100% | $\pm$ 2.0% |
| Skin Temperature | Exacon D-S18JK | 0 - 50 $°C$ | — |

In this study, the active approach is based on physiological information of occupants, such as height, weight, gender, skin temperature and its first-order derivative. The model developed is named as Predictive Thermal State (PTS) model. The model is also trained via neural networks with the features of skin temperature, its first-order derivative, normalized skin temperature and its normalized first-order derivative as shown in Figure 4. The normalization factor is based on the occupant's physiological parameters, like weight, height, gender, clothing, etc, and it is formulated as follows:

Equation 1

$$A_{du} = 0.203 \cdot Height^{0.725} \cdot Weight^{0.425}$$

$$A = (1 - I_{cl})A_{du}$$

$$T_{s_norm} = \frac{T_s}{A}, T'_{s_norm} = \frac{T'_s}{A}$$

## Problem Formulation

The problem is formulated as a minimization problem, and the objective function is defined with respect to energy consumption of electrical equipment and thermal comfort sensations of occupants, in which energy consumption and indoor thermal comfort are linearly weighted under a complementary relationship through a user-preference weight coefficient $\lambda$. The energy consumption and indoor thermal comfort levels are normalized beforehand, so that they are meaningful appearing in the objective function.

The objective function is defined as:

**Equation 2**

$$g(\lambda, \omega) = \lambda \cdot E_{norm}(\omega) + (1 - \lambda) \cdot |PTS|_{norm}(\omega)$$

The normalized energy consumption is defined as:

**Equation 3**

$$E_{norm}(\omega) = \frac{E(\omega) - \mu_E}{z \cdot \sigma_E}$$

The normalized $|PTS|$ is defined as:

**Equation 4**

$$|PTS|_{norm}(\omega) = \frac{|PTS(\omega)| - \mu_{|PTS|}}{z \cdot \sigma_{|PTS|}}$$

The Energy Saving Rate (ESR) is defined as:

**Equation 5**

$$ESR(\omega) = \frac{E(\omega) - E_{bench}}{E_{bench}} \times 100\%$$

Remarks: $\mu_E$, $\mu_{|PTS|}$, $\sigma_E$ and $\sigma_{|PTS|}$ are the corresponding means and standard deviations of energy consumption and thermal comfort states. $E_{bench}$ is the benchmark energy consumption at median designed operating frequencies $\omega = [40, 40, 40]^T$ of ACMV systems.

The optimization problem is formulated as:

Equation 6

$$Minimum\text{: } \min_{\forall\omega}(g(\lambda,\omega))$$
$$s.t.\ 0 \le \lambda \le 1$$
$$30 \le \omega \le 50$$

In the next sub-section, two methodologies of specific optimization algorithms are explained in detail.

## Bayesian Gaussian Process Optimization

Bayesian Gaussian Process optimization (BGPO) is based on the Bayesian probability with an assumption of Gaussian Process of observed samples. Gaussian Process (GP) is suitable for predicting outputs only based on few observed samples without knowing the exact function for difficult data-acquiring problem. Since BGPO method has these advantageous characteristics, it is selected for comparing with a non-Bayesian optimization method (i.e. Augmented Firefly Algorithm in this case). The definition and derivations of GP are presented below.

A Prior $P$ on a function $g(\cdot)$ is a Gaussian Process prior, with mean function $\mu_0$ and covariance function $k_0$. Thus, for any given set of $j$ observed samples, $\Omega = \{\omega_1, \omega_2, \dots \dots, \omega_j\}$ under $P$, we have

Equation 7

$$\begin{bmatrix} g(\omega_1) \\ g(\omega_2) \\ \vdots \\ g(\omega_j) \end{bmatrix} \sim \mathcal{N}\left(\begin{bmatrix} \mu_0(\omega_1) \\ \mu_0(\omega_2) \\ \vdots \\ \mu_0(\omega_j) \end{bmatrix}, \begin{bmatrix} k_0(\omega_1,\omega_1) & k_0(\omega_1,\omega_2) & \dots & k_0(\omega_1,\omega_j) \\ k_0(\omega_2,\omega_1) & k_0(\omega_2,\omega_2) & \cdots & k_0(\omega_2,\omega_j) \\ \vdots & \vdots & \ddots & \vdots \\ k_0(\omega_j,\omega_1) & k_0(\omega_j,\omega_2) & \cdots & k_0(\omega_j,\omega_j) \end{bmatrix}\right)$$

, where mean function is defined as follows:

Equation 8

$$\begin{bmatrix} \mu_0(\omega_1) \\ \mu_0(\omega_2) \\ \vdots \\ \mu_0(\omega_j) \end{bmatrix} = \begin{bmatrix} 0 \\ 0 \\ \vdots \\ 0 \end{bmatrix} = [0]$$

, where covariance (kernel) function is defined as follows:

Equation 9

$$k_0(\omega_i, \omega_j) = e^{\left(-\frac{1}{2\theta_h}\|\omega_i - \omega_j\|^2\right)}$$

Let

**Equation 10**

$$[K] = \begin{bmatrix} k_0(\omega_1, \omega_1) & k_0(\omega_1, \omega_2) & \cdots & k_0(\omega_1, \omega_j) \\ k_0(\omega_2, \omega_1) & k_0(\omega_2, \omega_2) & \cdots & k_0(\omega_2, \omega_j) \\ \vdots & \vdots & \ddots & \vdots \\ k_0(\omega_j, \omega_1) & k_0(\omega_j, \omega_2) & \cdots & k_0(\omega_j, \omega_j) \end{bmatrix}$$

Thus, Equation 7 can be further simplified as follows:

**Equation 11**

$$[g] \sim \mathcal{N}([0], [K]), \text{ where } [g] = \begin{bmatrix} g(\omega_1) \\ g(\omega_2) \\ \vdots \\ g(\omega_j) \end{bmatrix}$$

Since Gaussian distribution applies to all in the definition Domain, any given $\omega^*$ satisfies that

**Equation 12**

$$\begin{bmatrix} [g] \\ g(\omega^*) \end{bmatrix} \sim \mathcal{N}\left(\begin{bmatrix} [0] \\ \mu^* \end{bmatrix}, \begin{bmatrix} [K] & [k] \\ [k^T] & [k^*] \end{bmatrix}\right), \text{ where } [k] = \begin{bmatrix} k_0(\omega_1, \omega^*) \\ k_0(\omega_2, \omega^*) \\ \vdots \\ k_0(\omega_j, \omega^*) \end{bmatrix}, k^* = k_0(\omega^*, \omega^*)$$

Given Equation 12 the solution $g(\omega^*)$ follows a Gaussian distribution using Sherman-Morrison-Woodbury formula [32] as follows:

**Equation 13**

$$g(\omega^*) \sim \mathcal{N}(\mu^*, \Sigma^*), \text{ where } \mu^* = [k^T] \cdot [K^{-1}] \cdot [g], \Sigma^* = k^* - [k^T] \cdot [K^{-1}] \cdot [k]$$

Based on Equation 13, the optimal solution of $g(\omega)$ can be determined. After evaluations of different $\omega$ across the whole definition Domain, the best capable estimation can be achieved. The pseudo-code of BGPO is presented in Figure 5. The experimental parameters are shown in Table 3. The experiments were carried out under different sizes of samples from 10 to 50. Due to the constraints of operating conditions of ACMV systems, the boundaries of operating conditions are set as 50 and 30 for upper and lower boundary respectively.

Table 3: Experimental Parameters of Bayesian Gaussian Process Optimization

| Parameter | Sample | Boundary | $\theta_h$ |
|---|---|---|---|
| Value | 10/20/30/40/50 | 50/30 | 1 |

Observed initial samples $(\boldsymbol{\omega}, g(\boldsymbol{\omega}))$
While stopping criteria not satisfied
a) Calculate Bayesian posterior distribution from the observed samples $(\boldsymbol{\omega}, g(\boldsymbol{\omega}))$
b) Use posterior distribution to estimate next observation
Based on the most recent posterior distribution, report the point with the best estimation $(\boldsymbol{\omega}^*, g(\boldsymbol{\omega}^*))$.

Figure 5: BGPO Pseudo-code

## Augmented Firefly Algorithm Optimization

Augmented Firefly Algorithm (AFA) optimization is based on computational weight and distance learnings according to randomly distributed population. The pseudo-code is illustrated in Figure 6. The experimental parameters are presented in Table 4. The experiments were carried out under different sizes of samples ranging from 10 to 50. Due to the constraints of operating conditions of ACMV systems, the boundaries of operating conditions are set as 50 and 30 for the upper and lower boundaries. Gaussian distributions of randomness are selected due to better performances of earlier studies [14].

Table 4: Experimental Parameters of Sparse Augmented Firefly Algorithm

| Parameter | Sample | Boundary | $\alpha$ | $\beta$ | $\gamma$ | $\epsilon$ |
|---|---|---|---|---|---|---|
| Value | 10/20/30/40/50 | 50/30 | 0.6 | 0.3 | 0.6 | Gaussian $(0, 1)$ |

Objective function, $g(\boldsymbol{\omega})$, where $\boldsymbol{\omega} = [\omega^{(1)}, \omega^{(2)}, \omega^{(3)}]^T$
Generate initial population $n$ fireflies, $\boldsymbol{\omega}_i (i = 1,2,\dots,n)$
Evaluate intensity of fireflies, $I_i = g(\boldsymbol{\omega}_i)$
*while* (not reach stopping criteria)
*for* $i = 1\ to\ n$
  *if* $(I_i < I_{max})$
    $\boldsymbol{\omega}_i{}^{new} = \boldsymbol{\omega}_i{}^{old} + \alpha \cdot \gamma \cdot (\boldsymbol{\omega}_{max} - \boldsymbol{\omega}_i{}^{old}) + \beta \cdot [(\Delta B - 1) \cdot s + 1] \cdot \varepsilon$
    Update new intensity
  *else*
    $\boldsymbol{\omega}_i{}^{new} = \boldsymbol{\omega}_i{}^{old} + \beta \cdot [(\Delta B - 1) \cdot s + 1] \cdot \varepsilon$
    Update new intensity
  *endif*
*endfor*
*endwhile*
Rank population fireflies and obtain global optimum, $g^*$
Post-process and visualization.

**Notes:**
$\alpha = (0,1]$, Distance Coefficient
$\beta = [0,1]$, Randomness Coefficient
$\gamma = [0,1]$, Vortex Coefficient
$\varepsilon =$ Uniform/Gaussian Distribution
$\Delta B$ = Maximum Boundary Difference

$$s = \begin{cases} 0, & (Small\ Region\ Wandering) \\ 1, & (Large\ Region\ Wandering) \end{cases}$$

Figure 6: AFA Pseudo-code

## Polynomial Regression

Based on the discrete optimal solutions obtained from Bayesian Gaussian Process and Augmented Firefly Algorithm optimizers, 3th-order polynomial regression functions are formulated and trained. Define optimal solutions at $\lambda_i$ as follows:

Equation 14

$$y_i = \{E_i, PMV_i, ESR_i\} | \lambda_i$$

Since the user-preference weight coefficients are defined as $0 \leq \lambda \leq 1$ with an interval step of 0.1, we obtain a set of $n = 11$ discrete optimal solutions as follows:

Equation 15

$$(\boldsymbol{\lambda}, \boldsymbol{y}) = \{(\lambda_i, y_i), i = 1, 2, 3, \dots 11\}$$

We define a parameterized 3rd-order polynomial regression function as follows:

Equation 16

$$\tilde{y}_i = \theta_1 + \theta_2\lambda_i + \theta_3\lambda_i^2 + \theta_1\lambda_i^3$$

The cost function (MSE) of this parameterized regression function is defined as follows:

Equation 17

$$C(\Theta) = \frac{1}{2n}\sum_{i=1}^{n}(y_i - \tilde{y}_i)^2$$

The minimization of the cost function can be achieved by batch gradient descent method with respect to parameterized variable, $\Theta = [\theta_1, \theta_2, \theta_3, \theta_4]$, with a rational learning rate $\eta$.

Equation 18

$$\frac{\partial C}{\partial \theta_1} = \frac{2}{n}\sum_{i=1}^{n}(y_i - \tilde{y}_i) \Longrightarrow \theta_1^{new} = \theta_1^{old} - \eta\frac{\partial C}{\partial \theta_1}$$
$$\frac{\partial C}{\partial \theta_2} = \frac{2}{n}\sum_{i=1}^{n}(y_i - \tilde{y}_i)\lambda_i \Longrightarrow \theta_2^{new} = \theta_2^{old} - \eta\frac{\partial C}{\partial \theta_2}$$
$$\frac{\partial C}{\partial \theta_3} = \frac{2}{n}\sum_{i=1}^{n}(y_i - \tilde{y}_i)\lambda_i^2 \Longrightarrow \theta_3^{new} = \theta_3^{old} - \eta\frac{\partial C}{\partial \theta_3}$$
$$\frac{\partial C}{\partial \theta_4} = \frac{2}{n}\sum_{i=1}^{n}(y_i - \tilde{y}_i)\lambda_i^3 \Longrightarrow \theta_4^{new} = \theta_4^{old} - \eta\frac{\partial C}{\partial \theta_4}$$

## Experimental Results and Discussion

According to the topologies of neural networks modeling, the comparisons between different training features with genders are illustrated in Figure 7. The training features (skin temperature, first-order gradient of skin temperature, normalized skin temperature and normalized first-order gradient of skin temperature) are selected as optimal training features for predictive thermal state model, due to the best performance (over 96% accuracy) in both male and female tests.

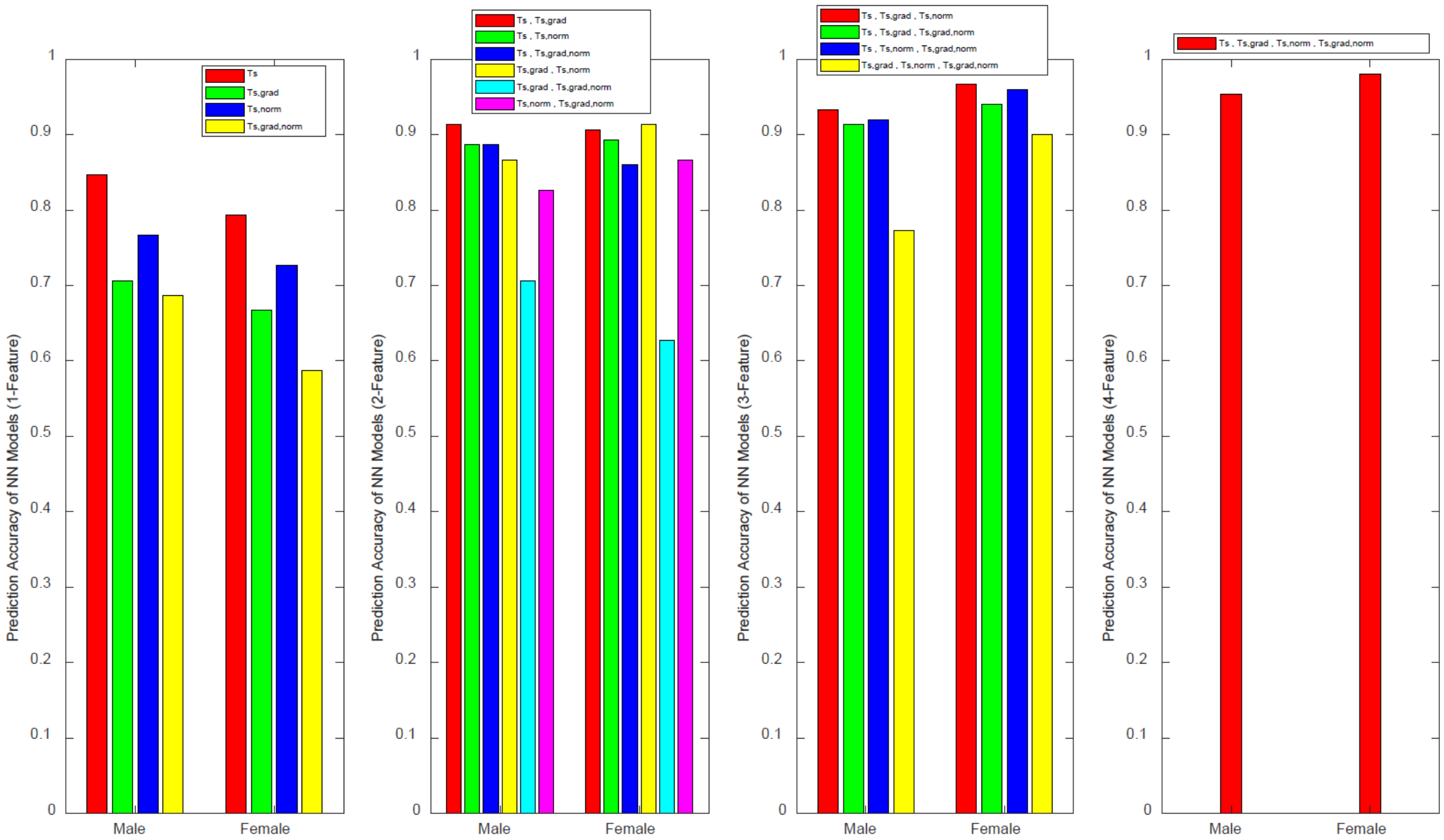

**Figure 7: Prediction Accuracy of PTS Modeling**

Similarly following the methodologies mentioned in the previous section, the experimental results of Bayesian Gaussian Process Optimization (BGPO) are shown in Figure 8 and 9, and the experimental results of Augmented Firefly Algorithm optimization (AFA) are shown in Figure 10 and 11. Both BGPO and AFA statistical analyses are tabulated in Table 5, 6 and 7. Based on the statistical analyses, there are two groups for each of the two cases, Case 1 and Case 2. The groups are classified by $\lambda$ values. The thermal comfort preferred (TCP) group is classified as $\lambda \leq 0.3$. Similarly, for $\lambda \geq 0.7$, the group is classified as energy-efficiency preferred (EEP) group. Comparing Table 5 and 6, BGPO generally outperforms AFA in terms of searching for more optimal solutions, since the means of Energy Saving Rate (ESR) are around -21% and -10% for BGPO and AFA respectively for both Case 1 and Case 2. However, AFA takes over BGPO in terms of consistency, and the standard deviations of ESR are around 0.02 and 0.006 for BGPO and AFA respectively. The increment of sample sizes improves searching in precision and consistency as presented in Table 5 and 6. As seen in Table 7, the evaluations at a sample size of 50 reveal that AFA is superior to BGPO in terms of consistency. Moreover, BGPO searches more optimal solutions of ESR than AFA on Case 1 and Case 2. However, AFA locates more precise solutions of PMV than BGPO on Case 1 and Case 2. Furthermore, the ESR and PMV results of Case 1 are more optimal than those of Case 2, due to different environmental resistance to ACMV systems.

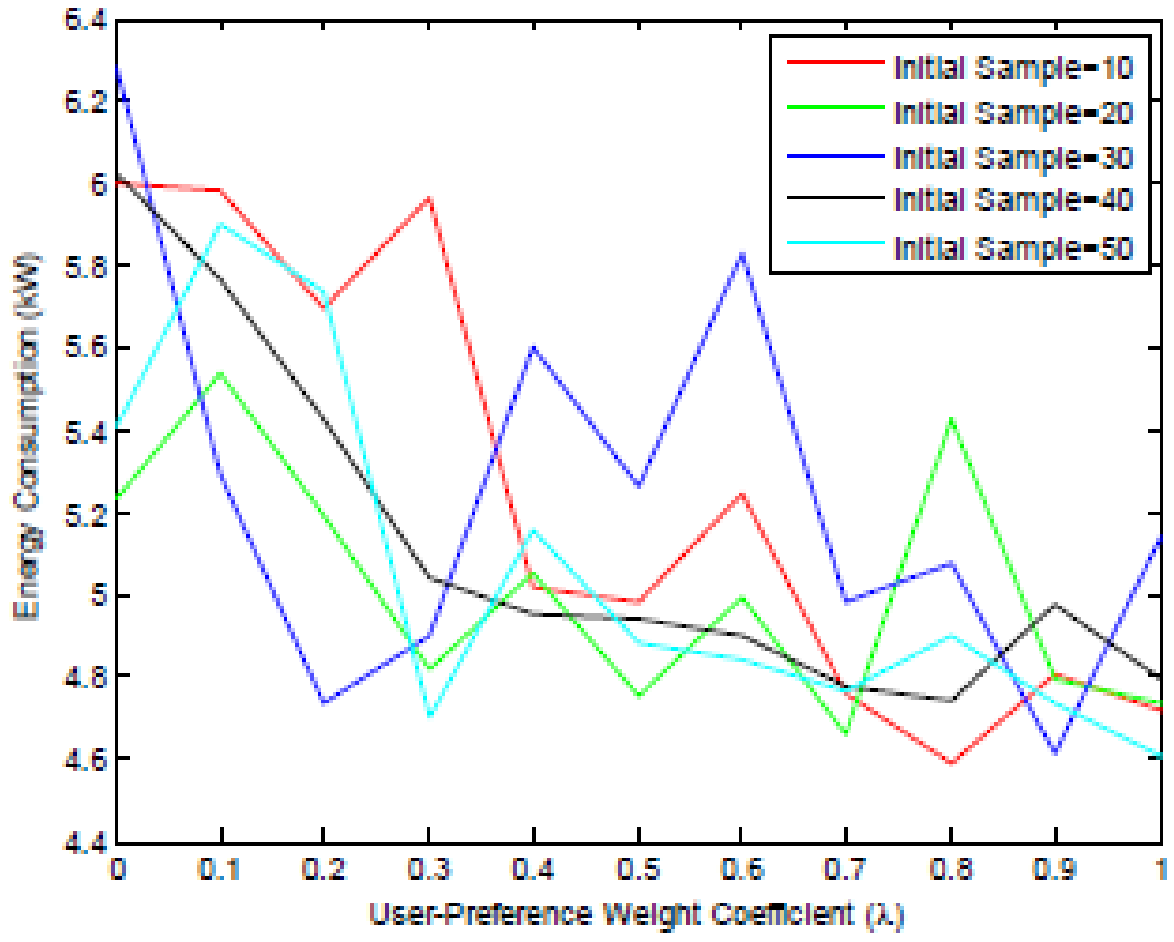

(a) Energy Consumption - Discrete

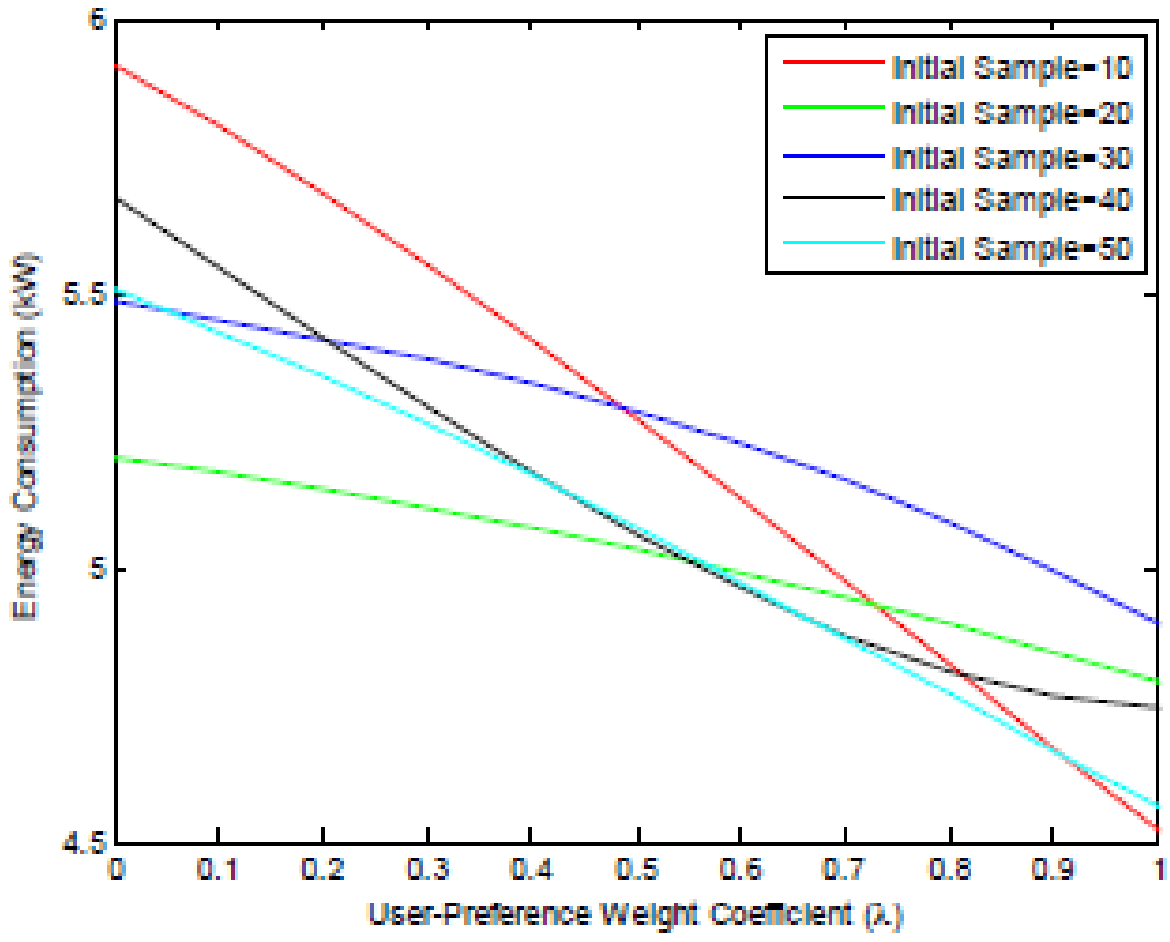

(b) Energy Consumption - Regression

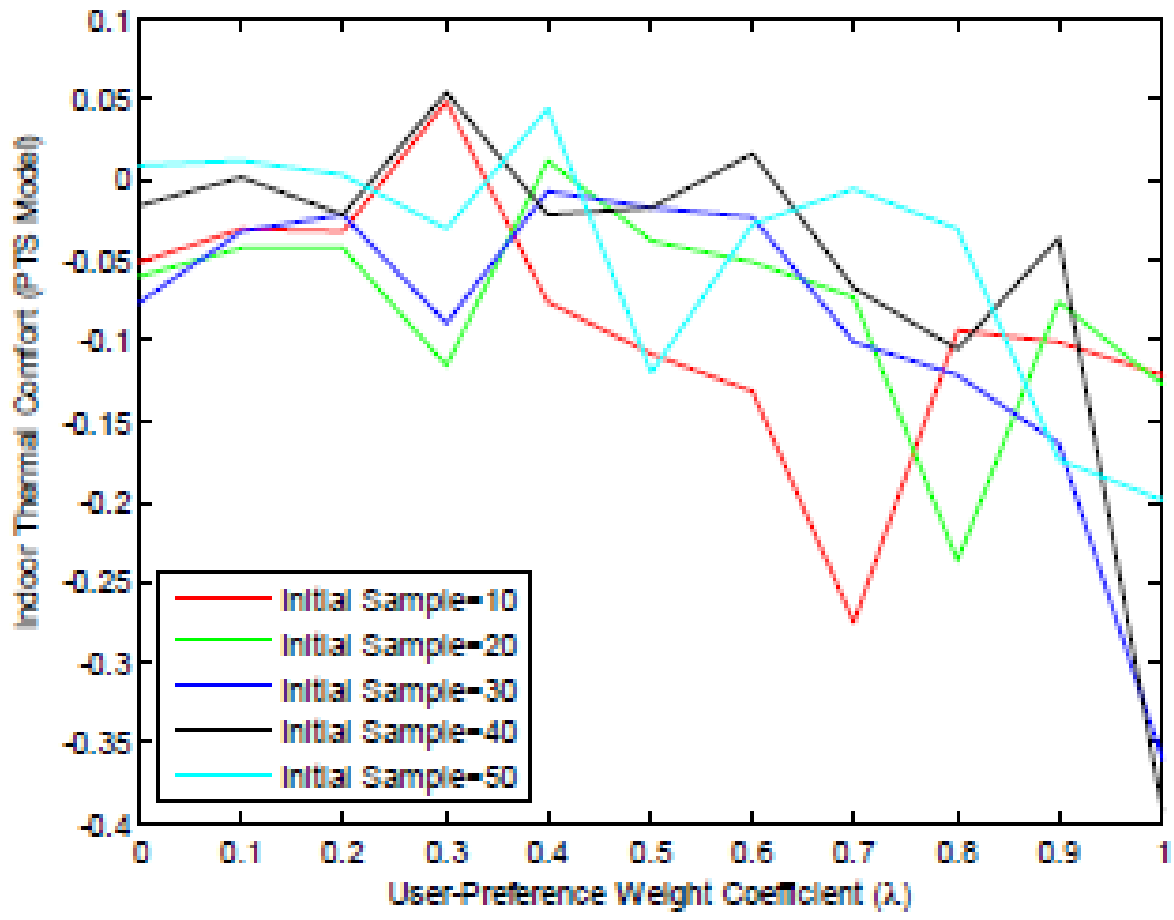

(c) Indoor Thermal Comfort - Discrete

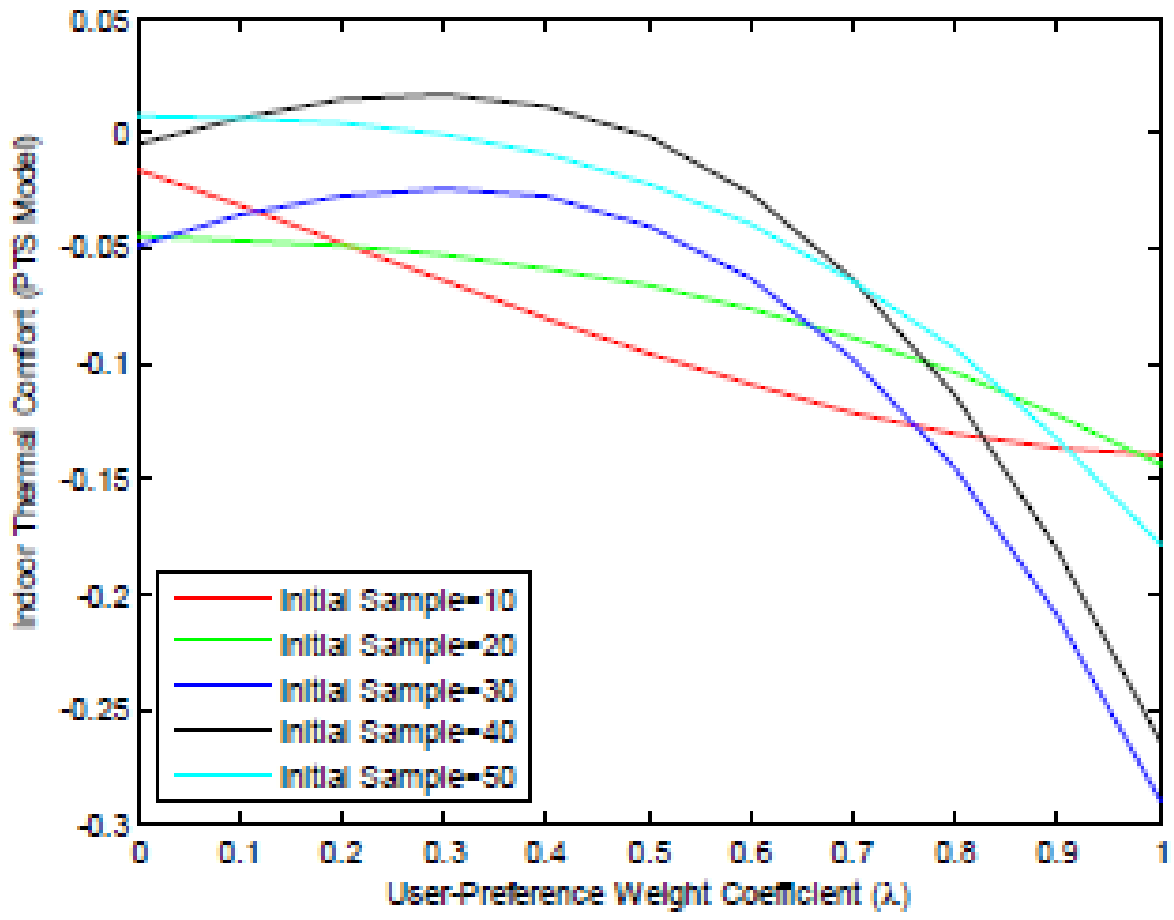

(d) Indoor Thermal Comfort - Regression

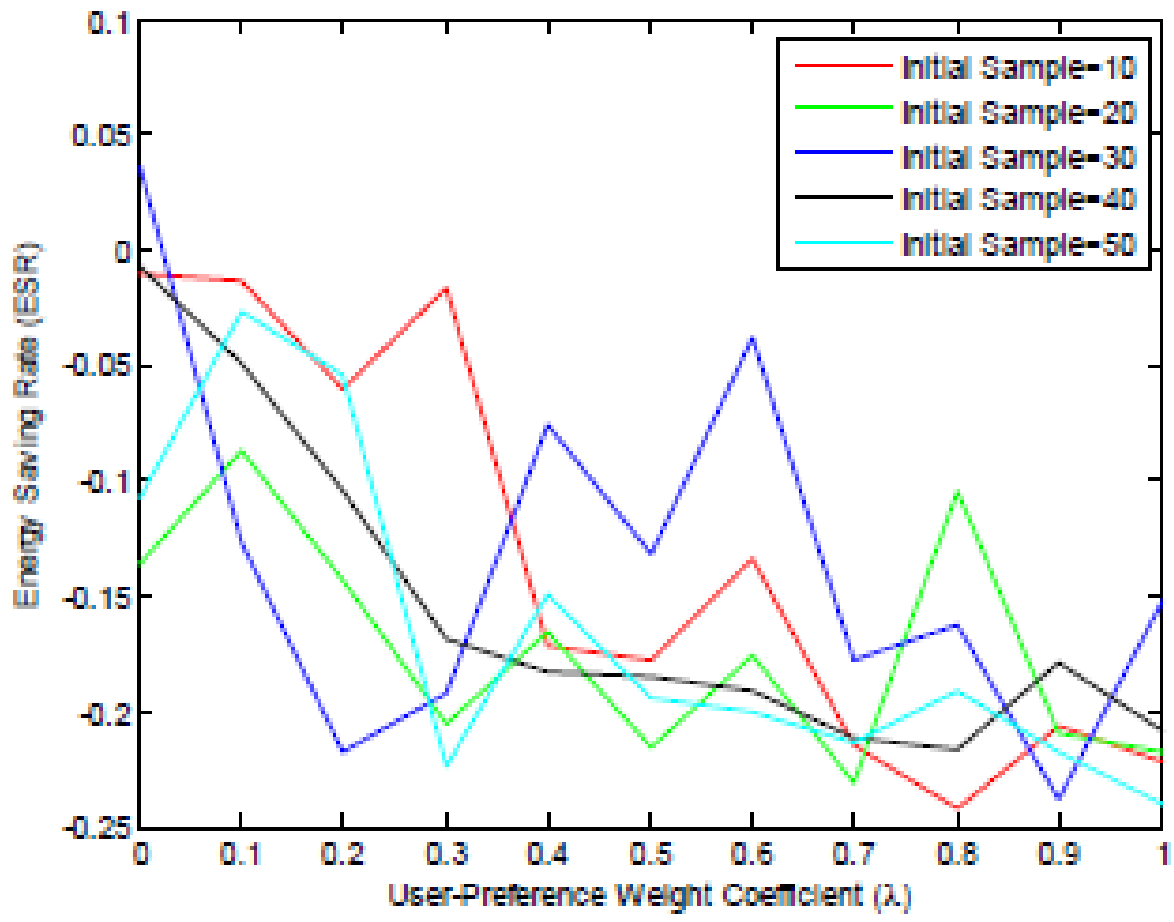

(e) Energy Saving Rate - Discrete

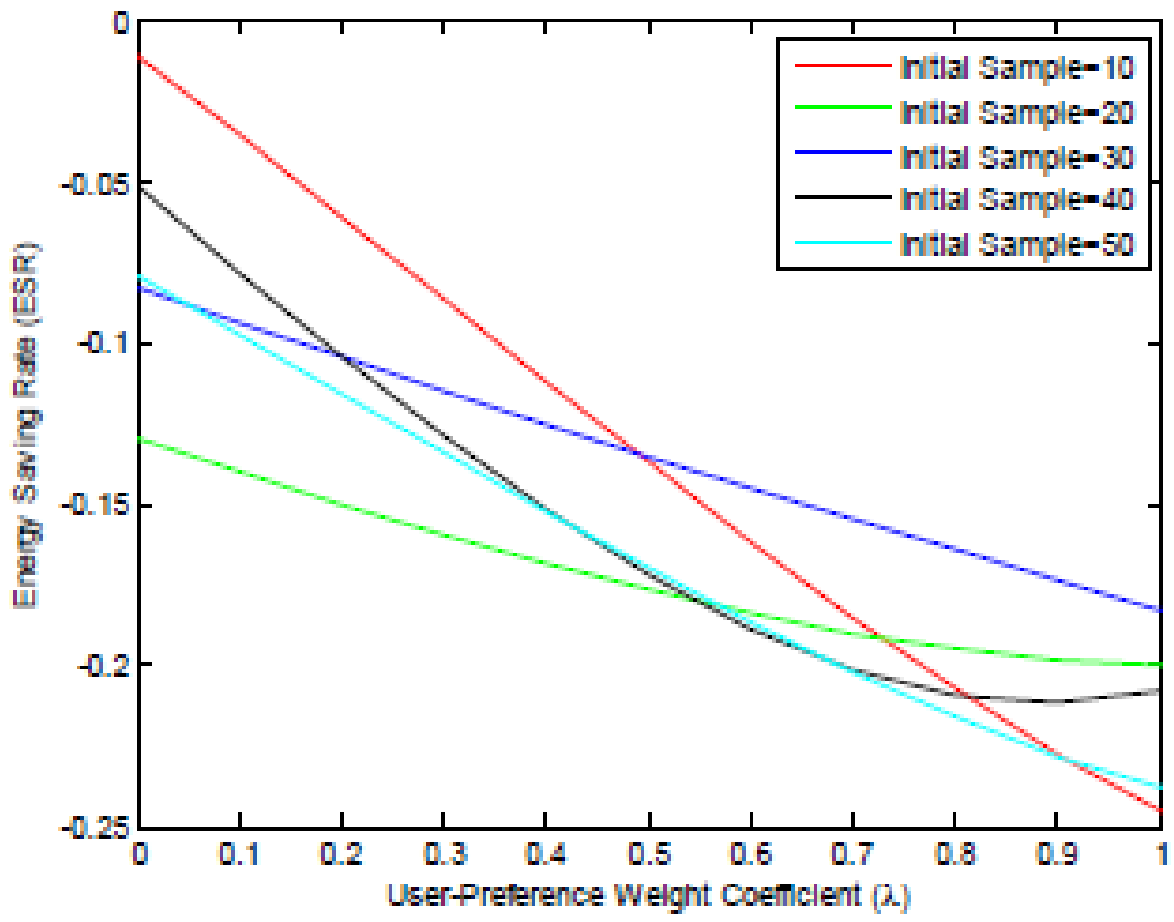

(f) Energy Saving Rate - Regression

**Figure 8: BGPO Discrete and Regression Results for Case 1**

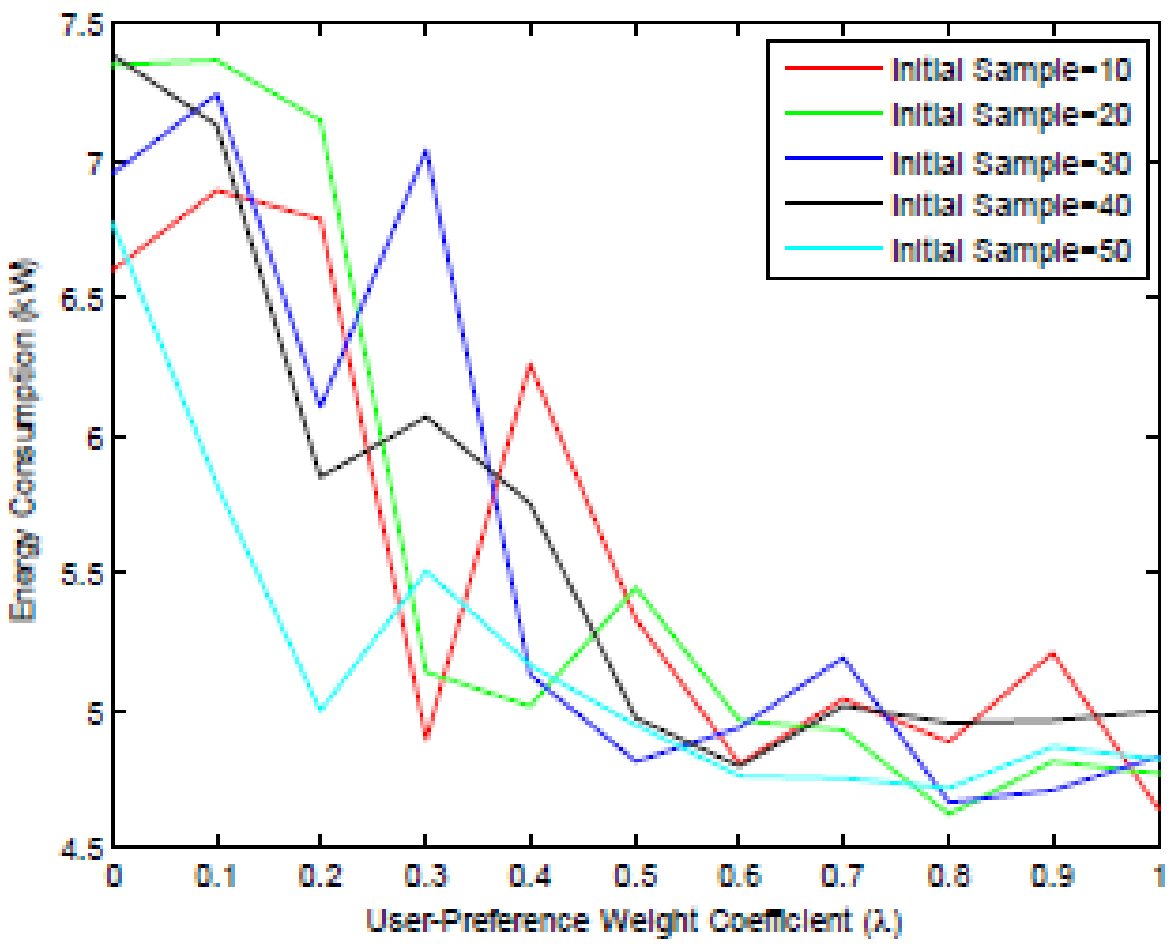

(a) Energy Consumption - Discrete

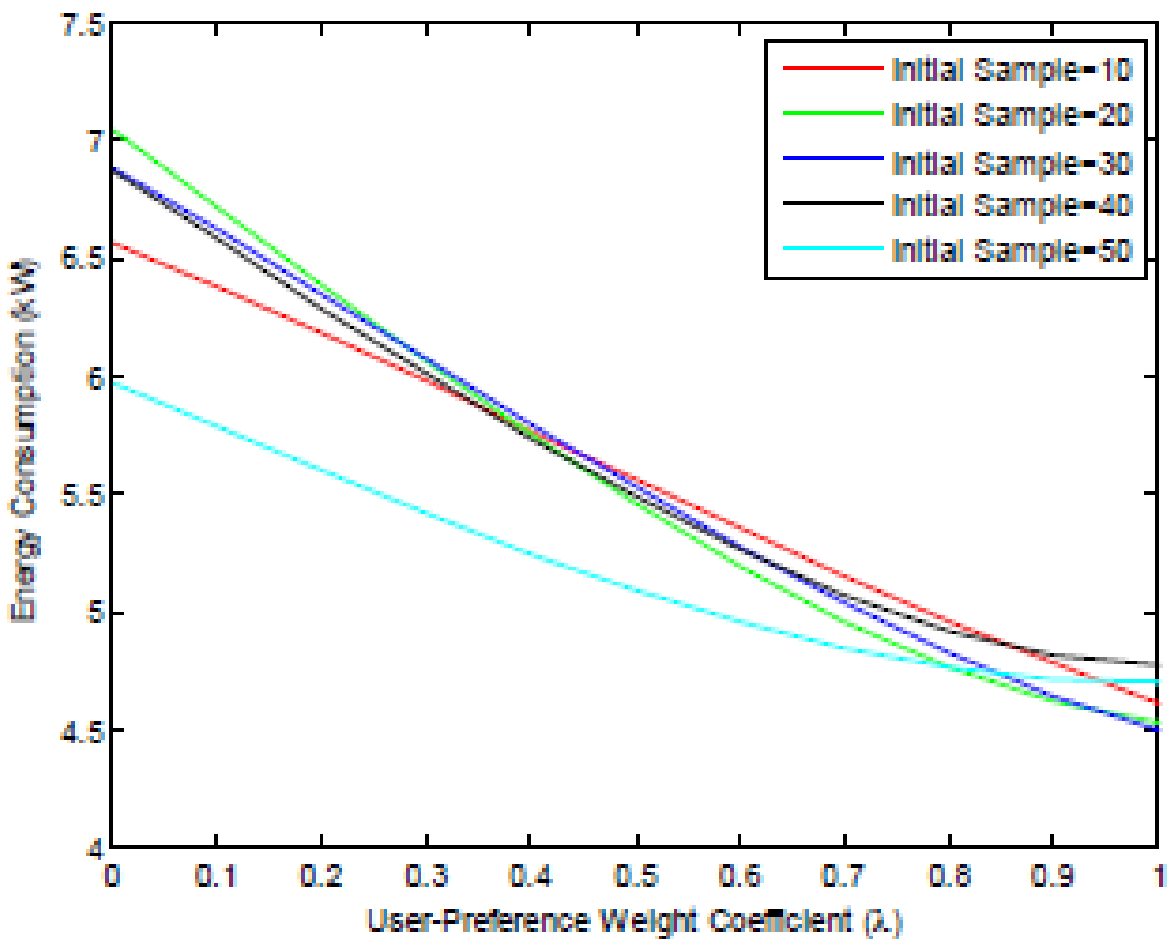

(b) Energy Consumption - Regression

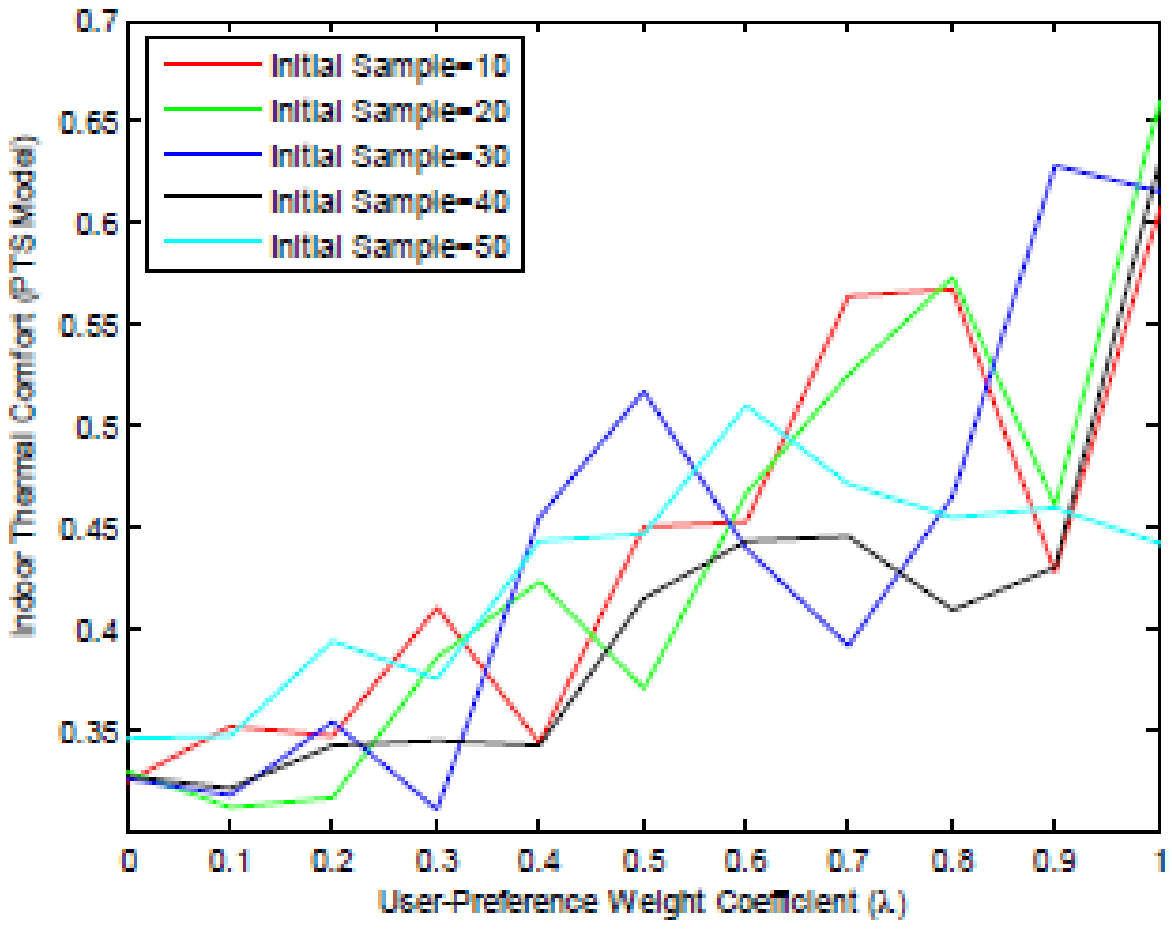

(c) Indoor Thermal Comfort - Discrete

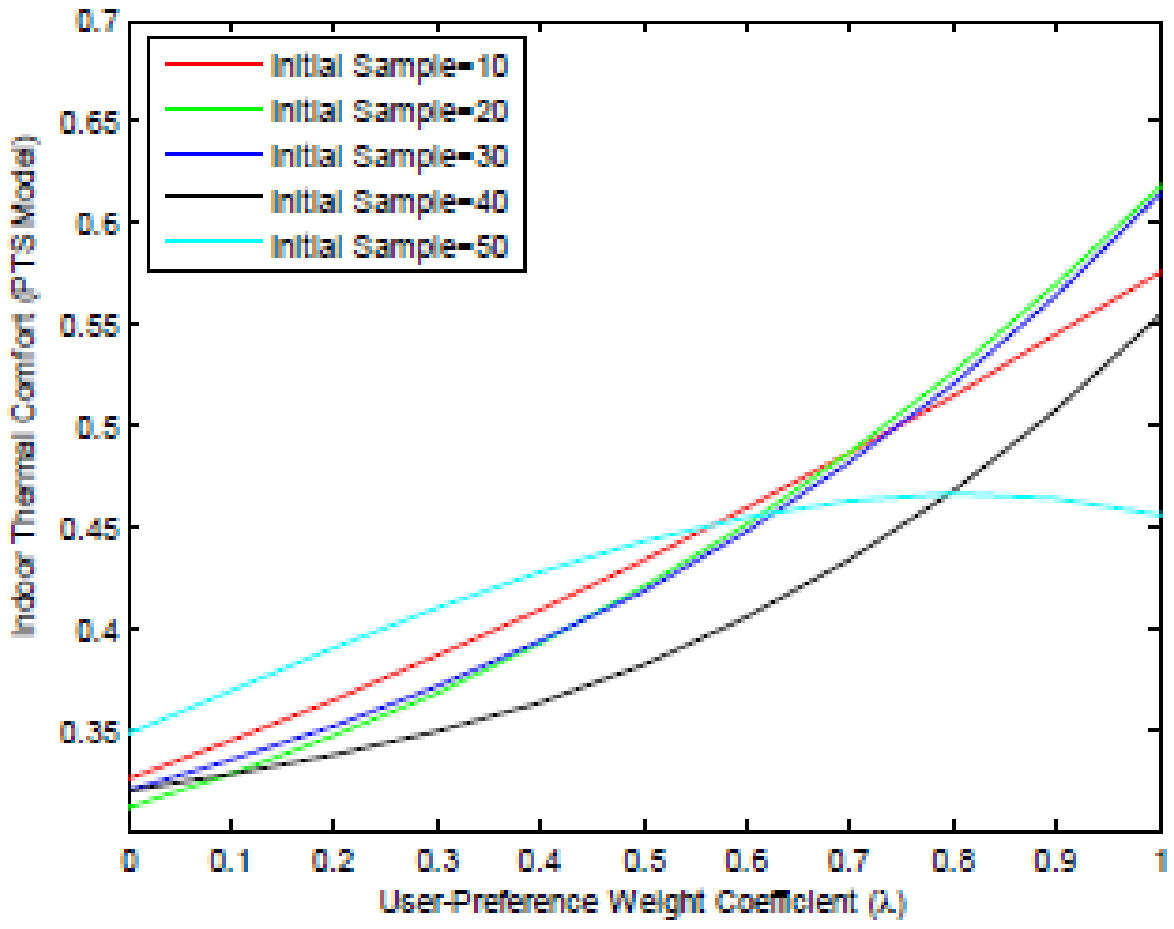

(d) Indoor Thermal Comfort - Regression

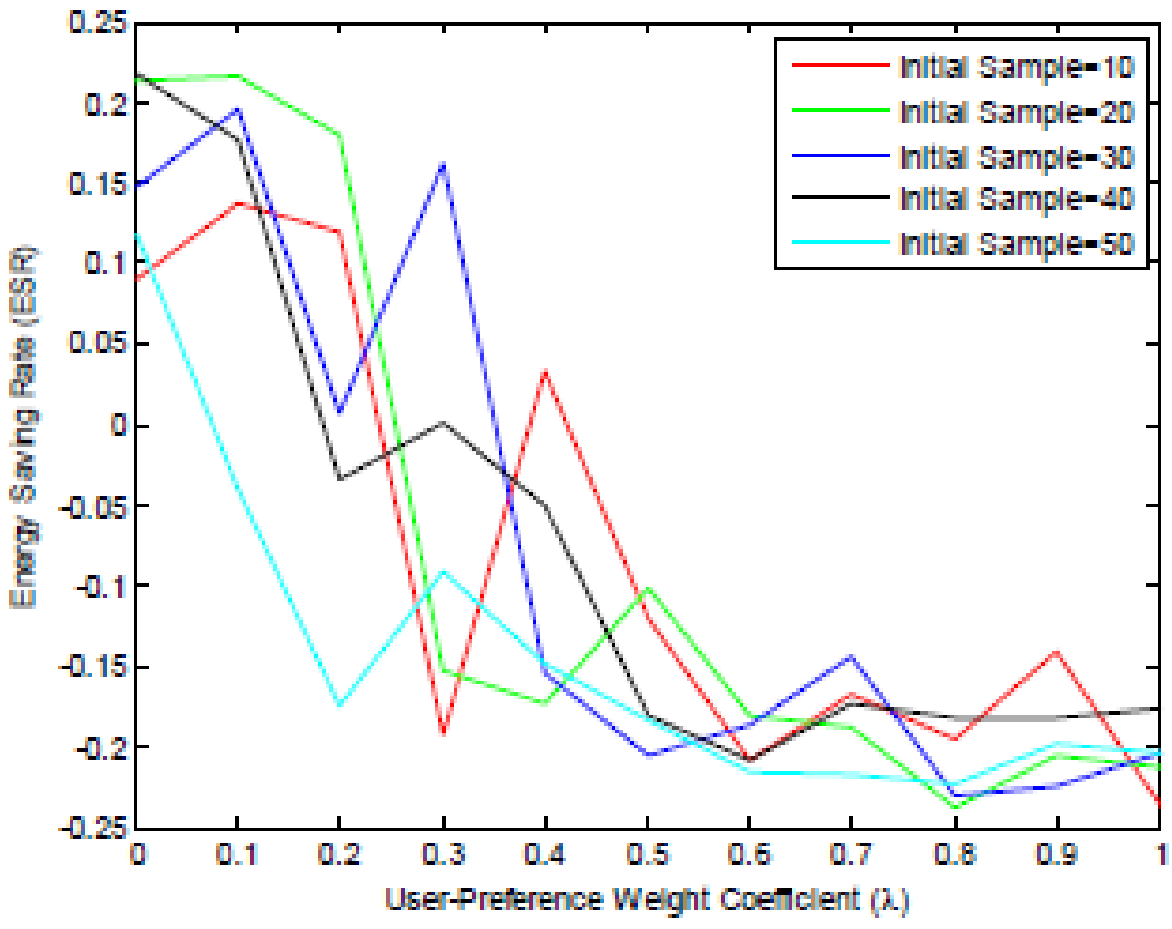

(e) Energy Saving Rate - Discrete

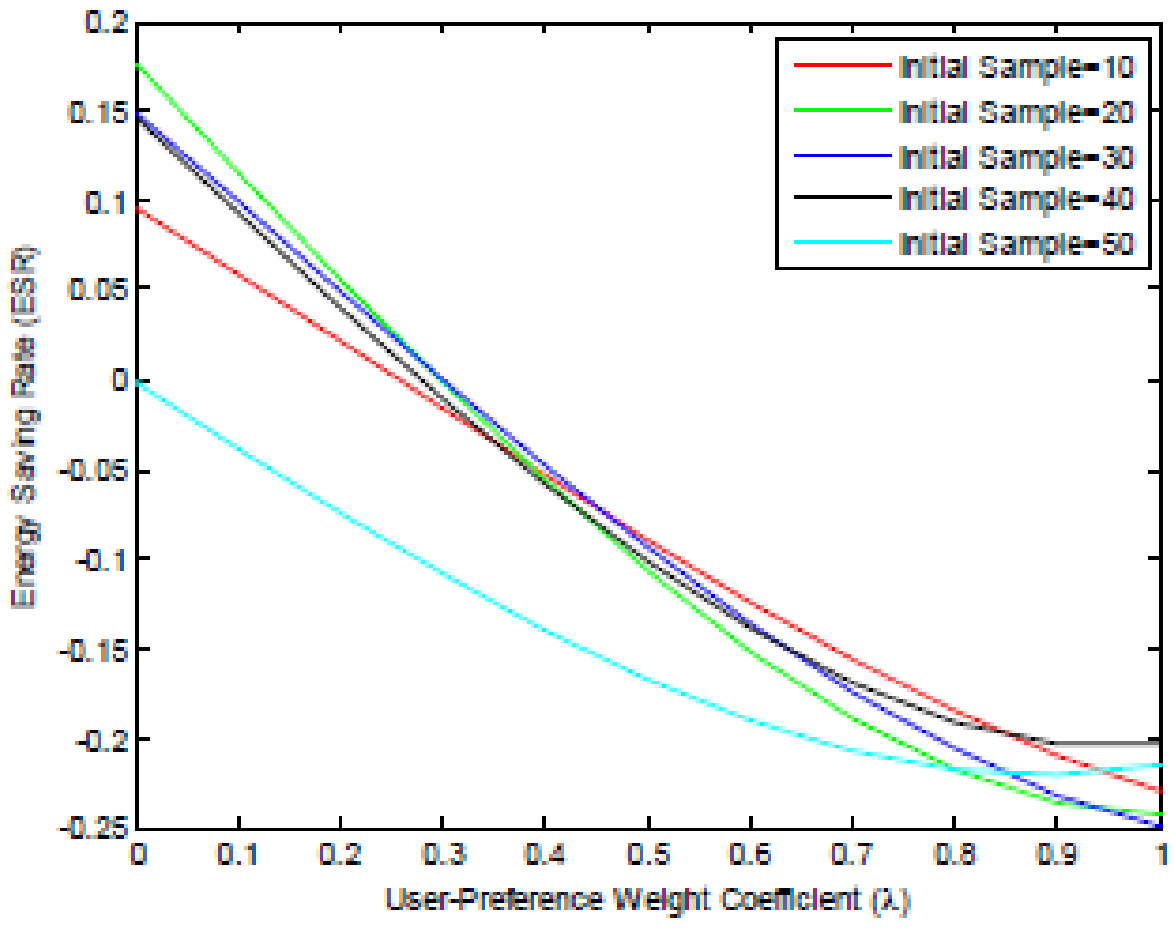

(f) Energy Saving Rate - Regression

**Figure 9: BGPO Discrete and Regression Results for Case 2**

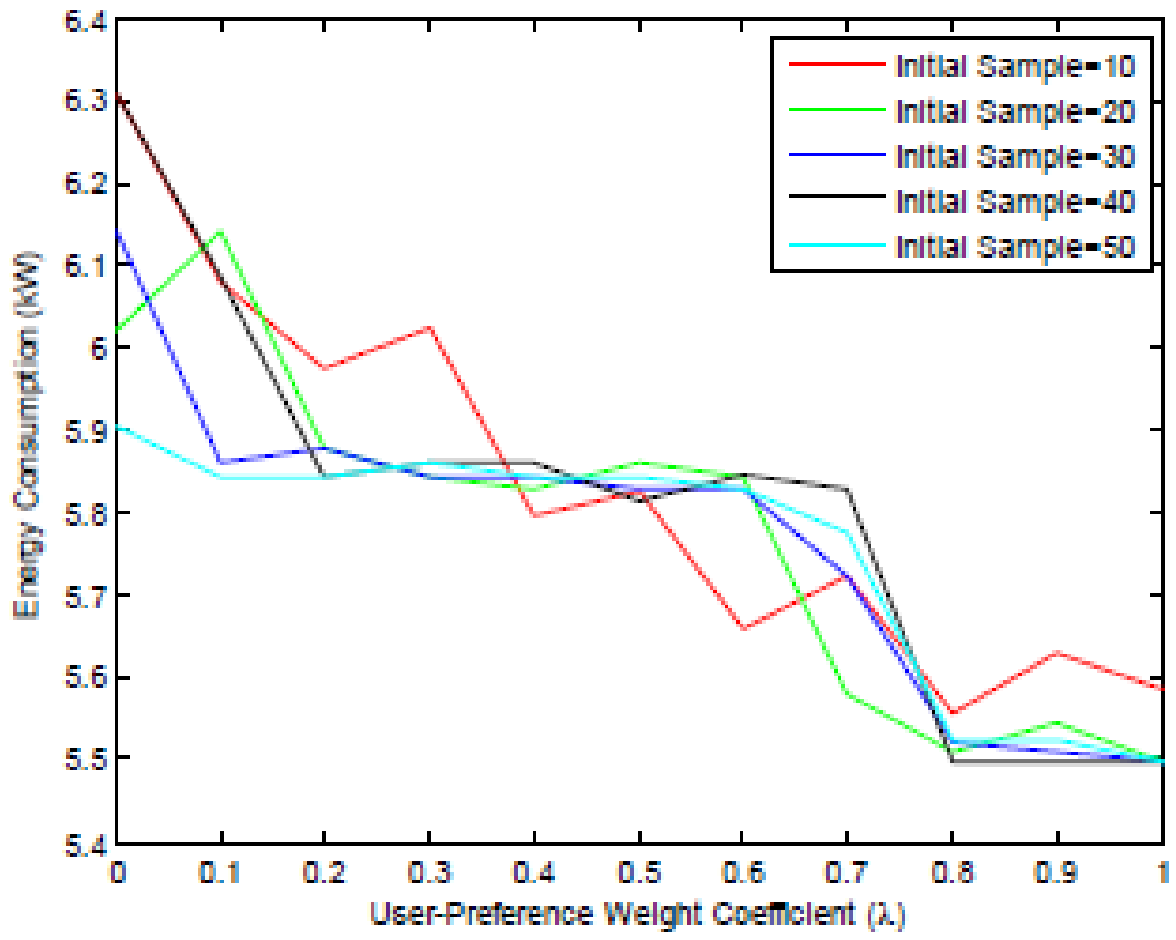

(a) Energy Consumption - Discrete

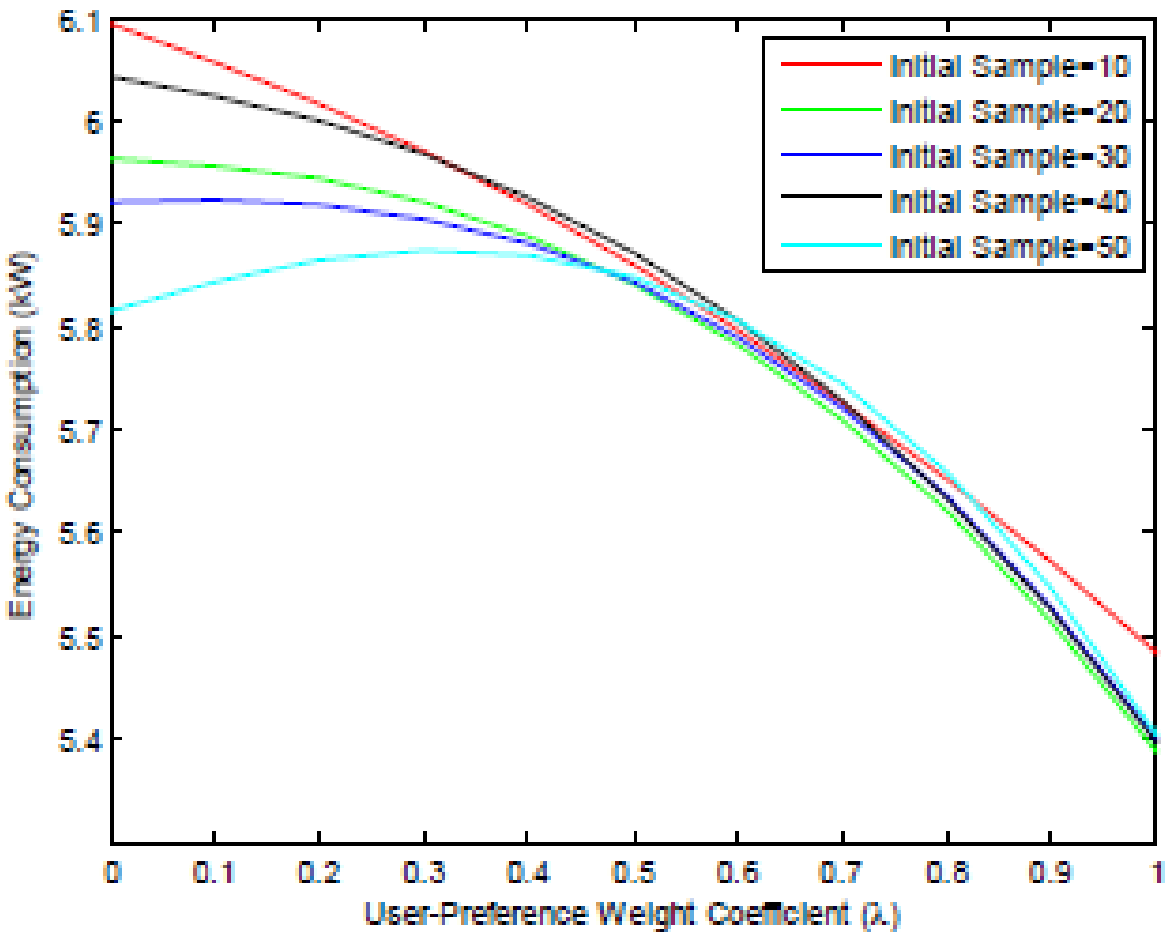

(b) Energy Consumption - Regression

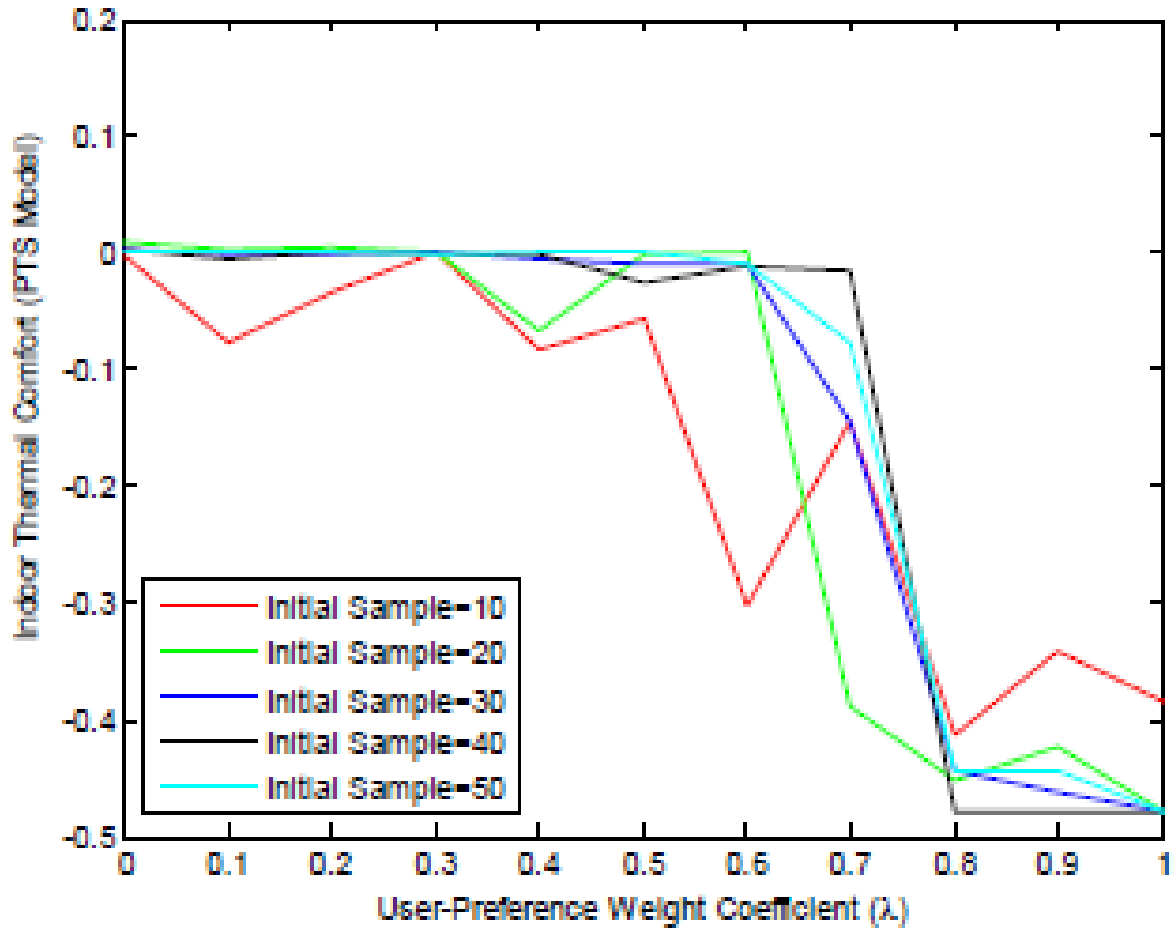

(c) Indoor Thermal Comfort - Discrete

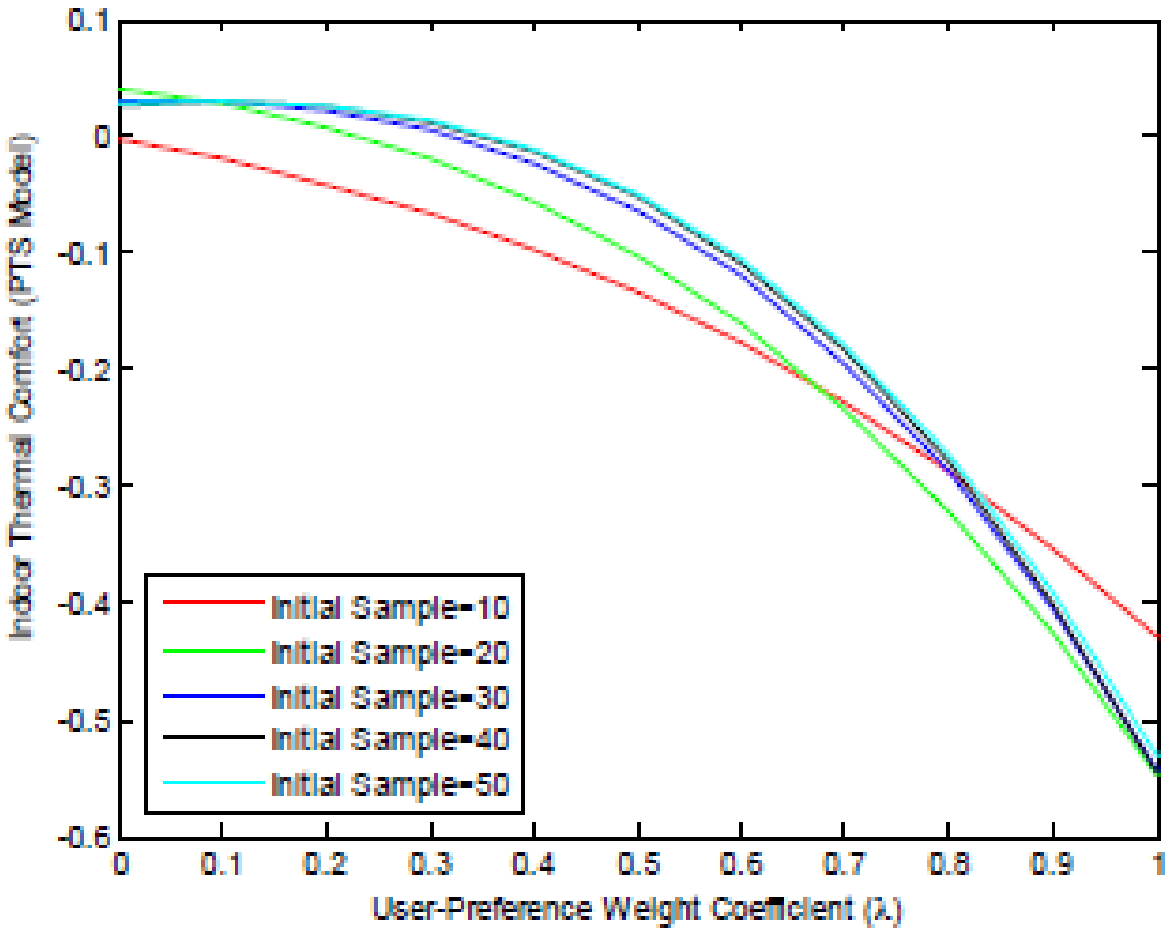

(d) Indoor Thermal Comfort - Regression

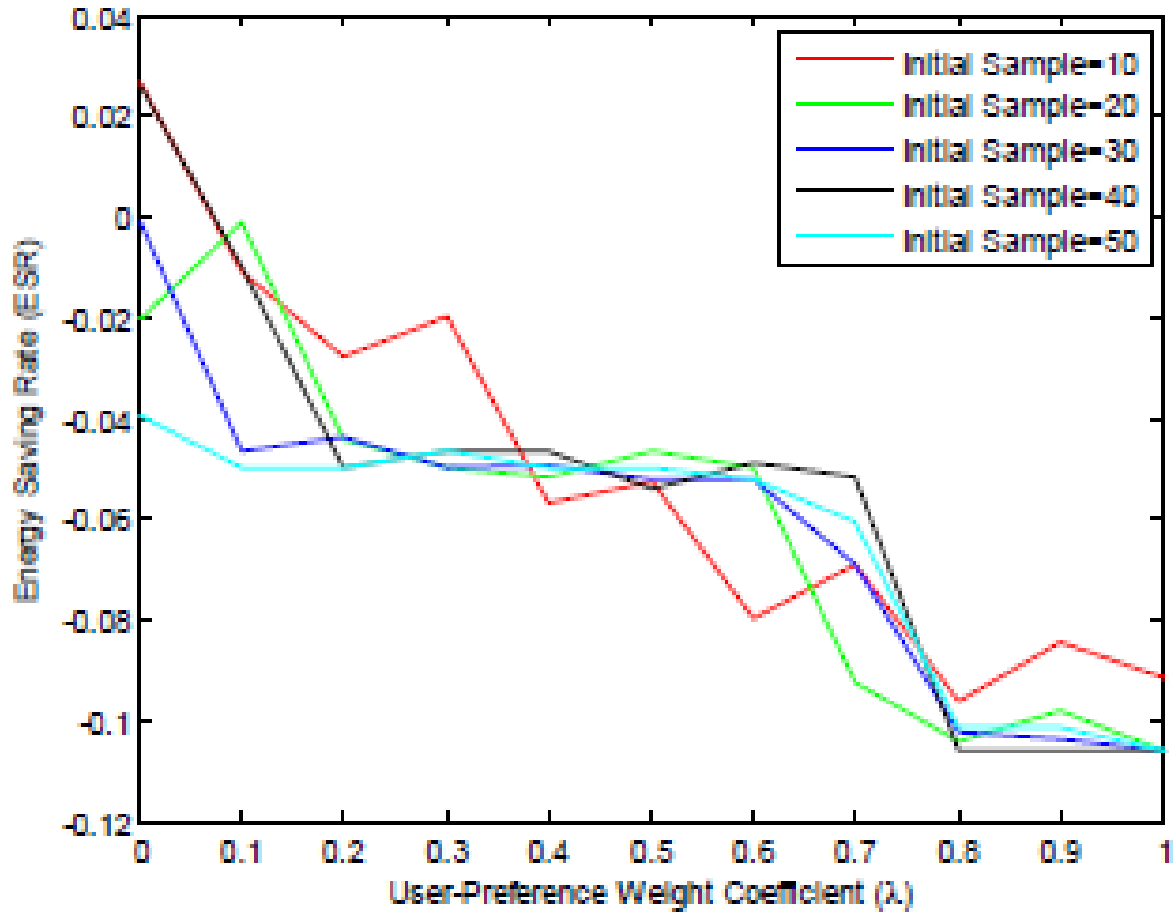

(e) Energy Saving Rate - Discrete

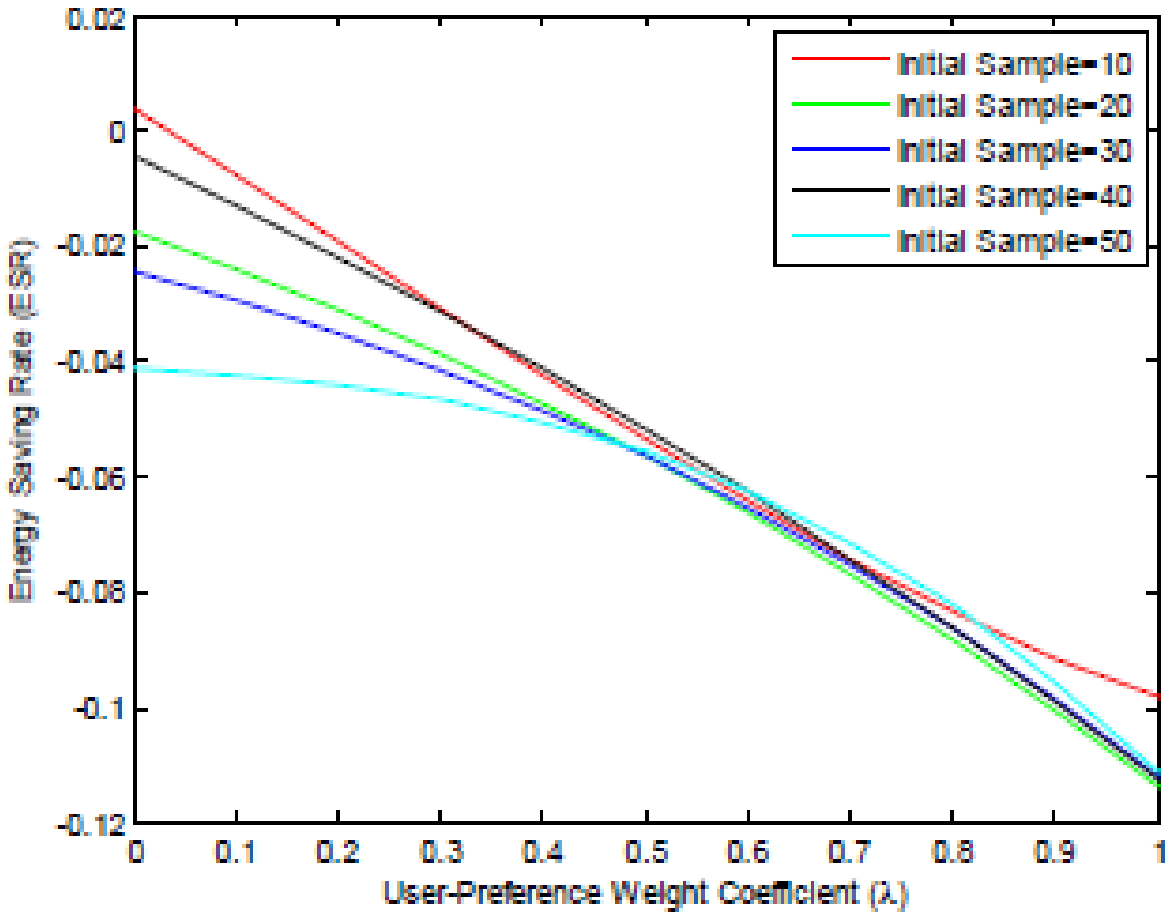

(f) Energy Saving Rate - Regression

**Figure 10: AFA Discrete and Regression Results for Case 1**

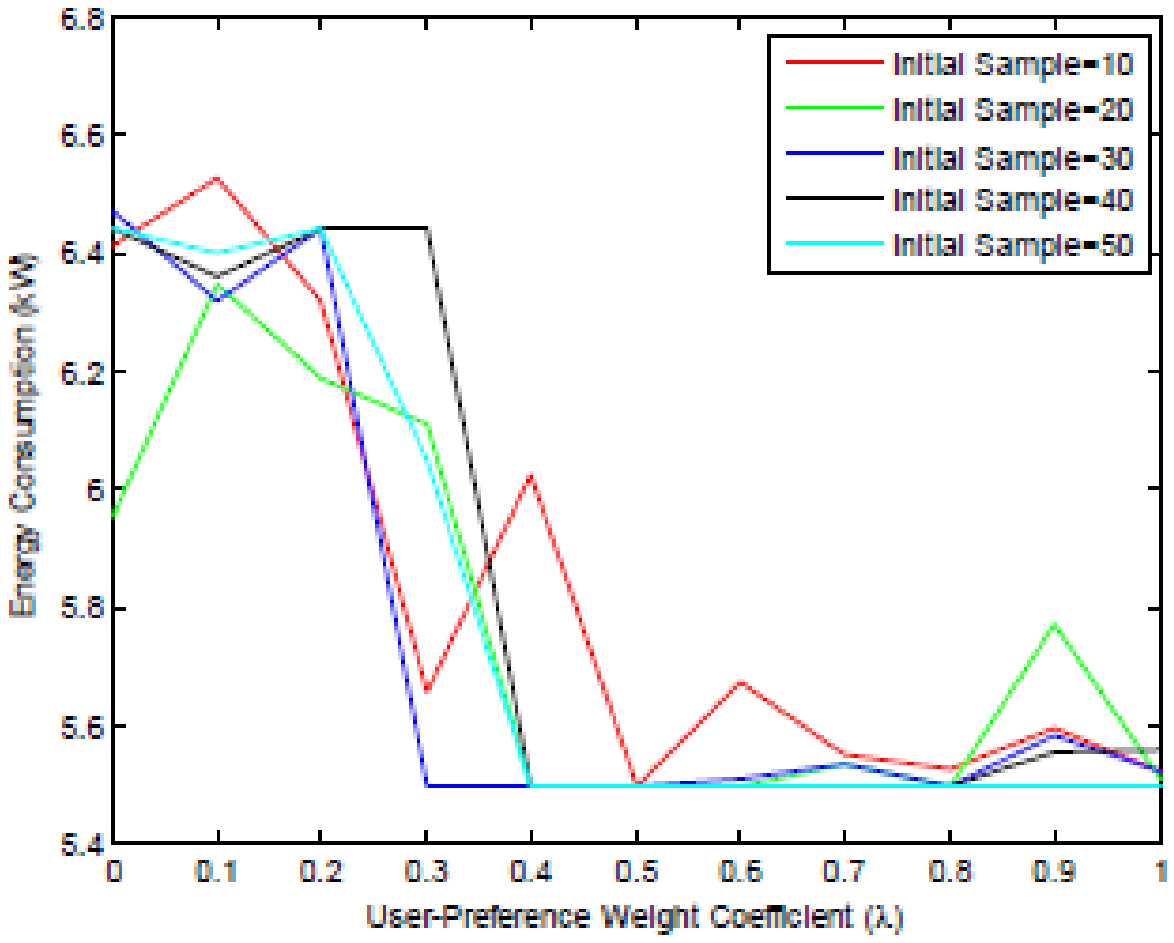

(a) Energy Consumption - Discrete

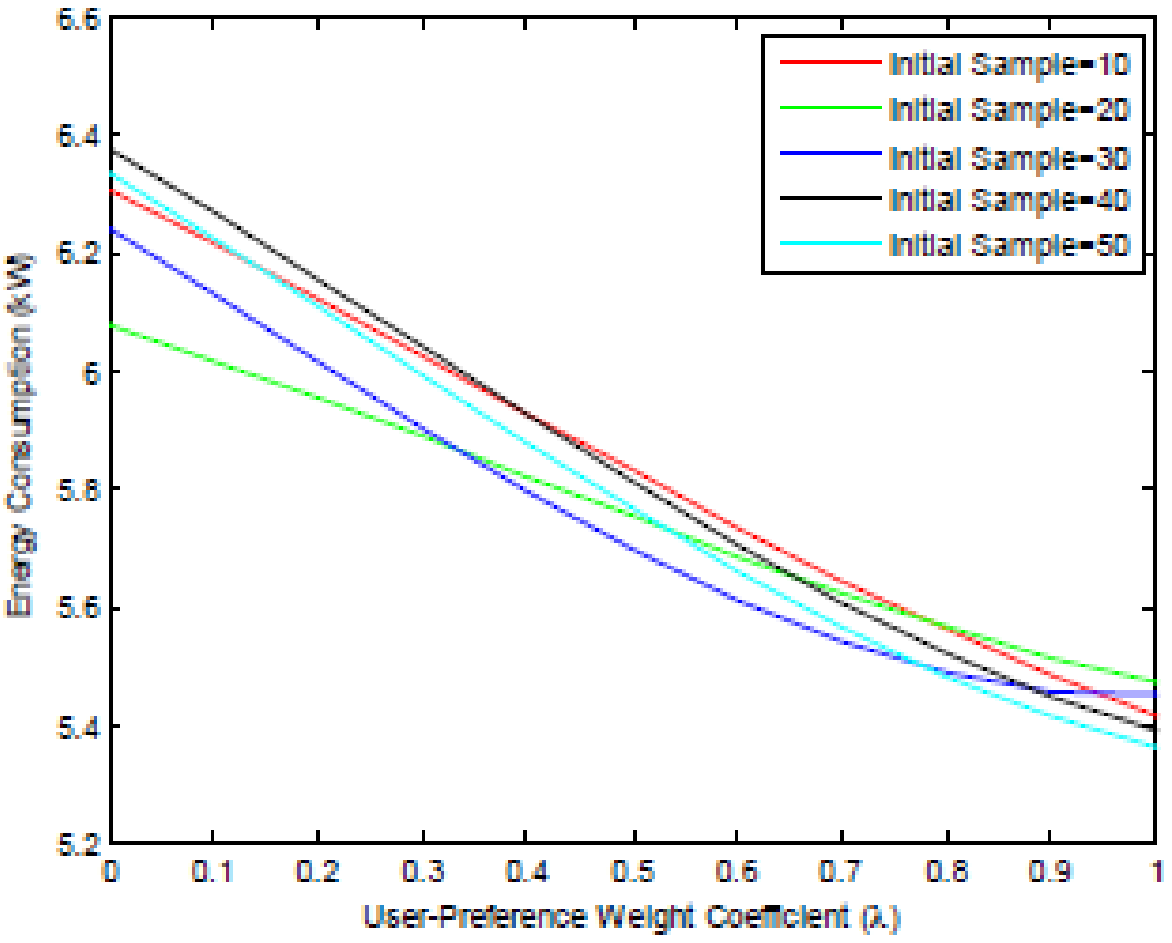

(b) Energy Consumption - Regression

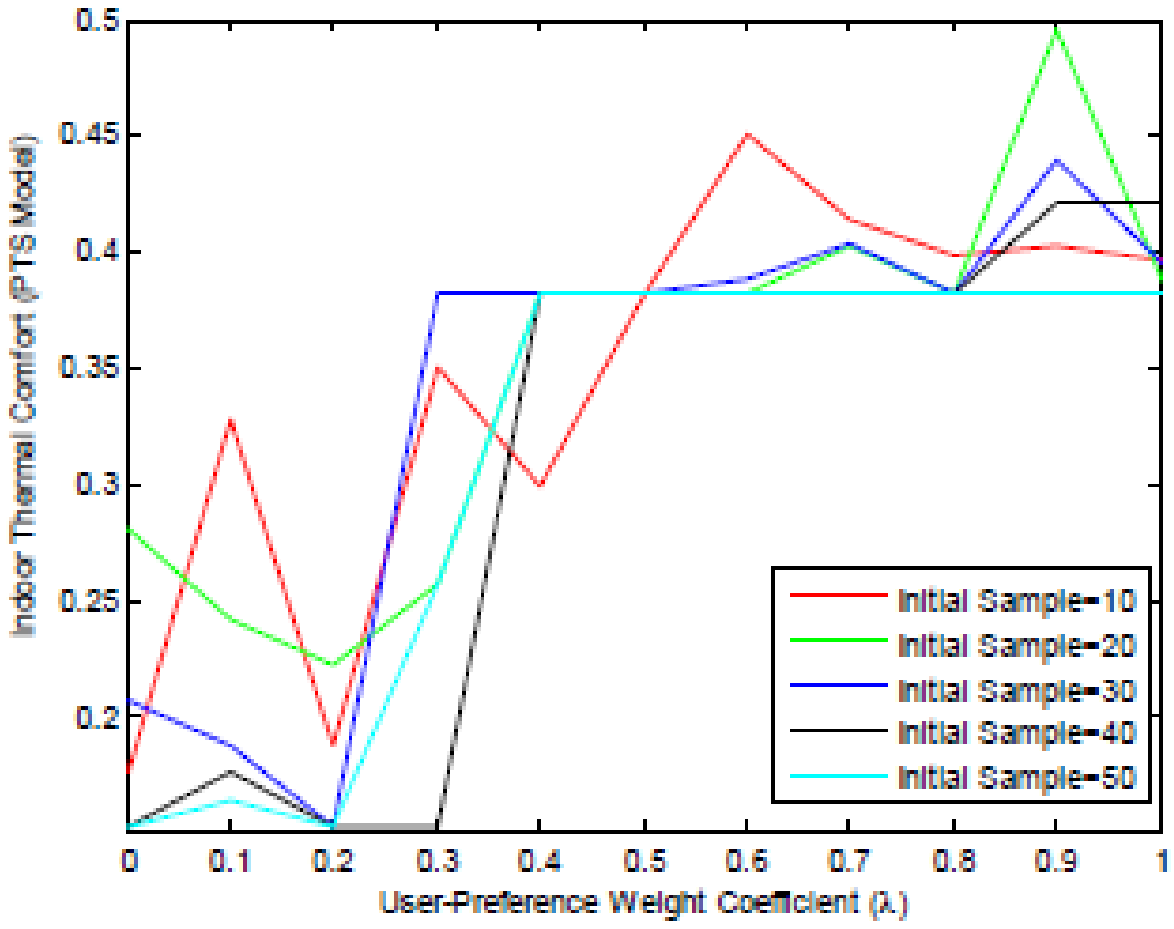

(c) Indoor Thermal Comfort - Discrete

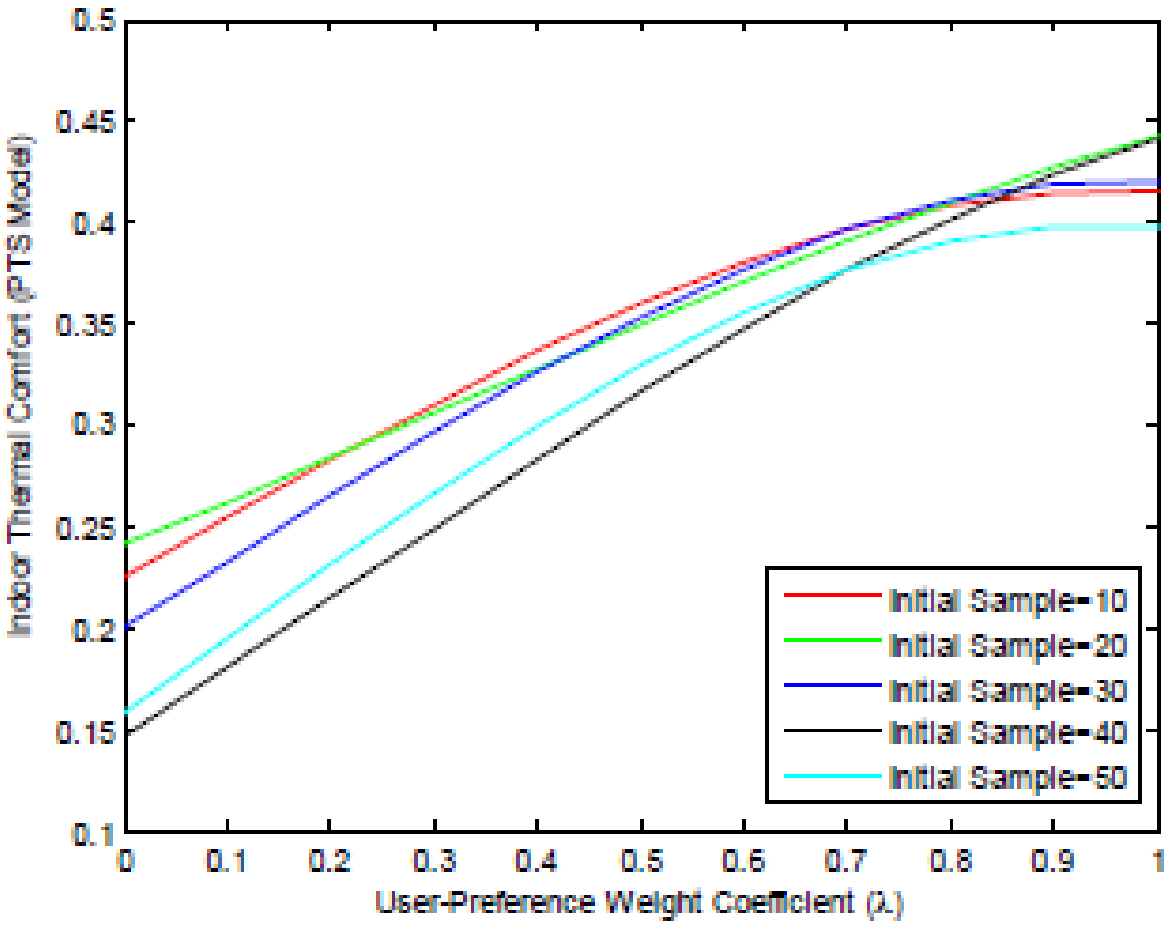

(d) Indoor Thermal Comfort - Regression

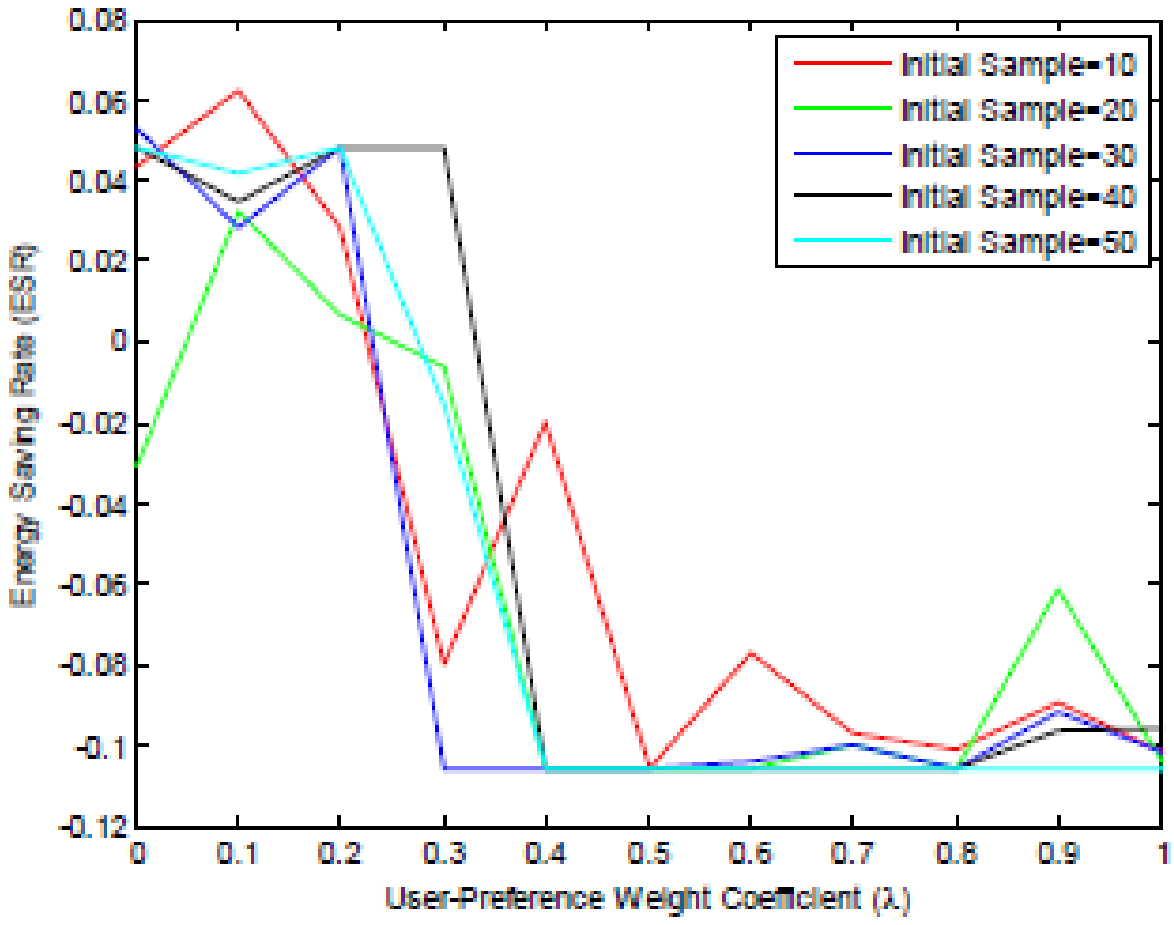

(e) Energy Saving Rate - Discrete

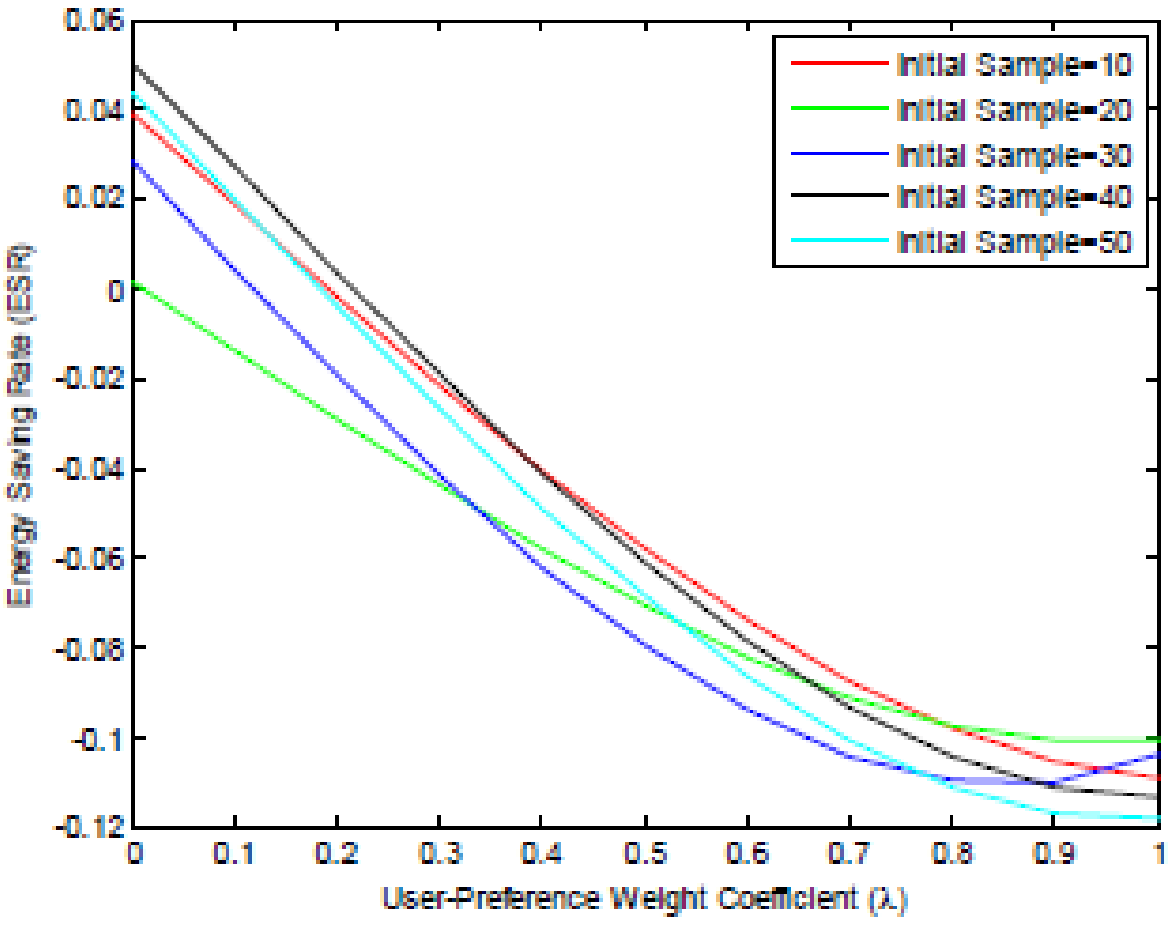

(f) Energy Saving Rate - Regression

**Figure 11: AFA Discrete and Regression Results for Case 2**

## Conclusion

Based on the experimental results and analyses of Bayesian Gaussian Process Optimization (BGPO) and Augmented Firefly Algorithm (AFA) above, we can draw the following conclusions.

First, BGPO outperforms AFA in terms of Energy Saving Rate (ESR) in EEP group. BGPO has an ESR of -21% for Case 1 and Case 2 in EEP group, while AFA has an ESR of -10% for Case 1 and Case 2 in EEP group. As the figures presented, BGPO can perform more energy-saving than AFA does on average. Generally, the experimental laboratory consumes 73.741 kWh per day for office hours. In terms of economical performances of these two methodologies, the Singapore Energy Market Authority announced that the electricity costs 21.56 cents per kWh, and it results in saving about S$1219.1 annually for such single experimental room based on BGPO methodology. If the whole rooms in the building were governed by the methodology, the electricity tariff could be significantly reduced and demonstrate a large number of economic benefits.

Second, AFA not only surpasses BGPO in terms of indoor thermal comfort (i.e. PMV index), but also in solution consistency in TCP group. The standard deviation of BGPO is around 0.02 for Case 1 and Case 2. The standard deviation of AFA is about 0.006 for Case 1 and Case 2. BGPO and AFA have PMV indices of 0.0087 and 0.0008 respectively for Case 1 with a sample size of 50. Similarly, BGPO and AFA have PMV indices at 0.3463 and 0.1524 respectively for Case 2 with a sample size of 50. As shown in the figures (Figure 8-11), although both of BGPO and AFA have located the PMV indices within the pre-defined comfort zone, AFA can locate around 10 times and 2 times better than BGPO does for Case 1 and Case 2 respectively.

Third, the improvements of ESR and PMV can be achieved by both BGPO and AFA with incremental sample sizes. The optimization results in Case 1 are slightly better than those in Case 2, due to different physiological parameters of the experimental occupant. The activity level and clothing factor in Case 1 are significantly lower than that in Case 2. Thus, the cooling demand in Case 2 is higher than that in Case 1, and the ACMV systems do more work to meet the cooling demand in Case 2.

In this study, there are several limitations that should be noted.

- The thermal laboratory is located in a tropical country (i.e. Singapore). Therefore, the ACMV systems only provide the cooling capacity, which is not a general Heating and Ventilation Air-Conditioning (HVAC) system for all four seasons.
- There is an assumption that the rooms in the buildings are identical for the purpose of experimental simplicity. However, in real applications, each room is different according to its facing, location inside the building and many other factors. Therefore, there is a need for room classifications if we aim to comprehensively improve the energy-efficiency of the whole building.
- The indoor comfort actually consist more than just thermal comfort, and there are luminous comfort and air quality as well. In this study, the lighting is always provided due to the situation of thermal laboratory inside the whole ACMV laboratory. In addition, since there is only one occupant in this thermal laboratory during experiments, the air quality (such as CO2) was always within the acceptable range while experiments were conducted. Therefore, luminous comfort and air quality of the experimental room are not in our scope of this study, and only indoor thermal comfort has been examined.

## Acknowledgment

The authors would like to express their many sincere thanks to the Center of EXQUISITUS for providing the experiment air-conditioning mechanical ventilation systems in the thermal laboratory of Nanyang Technological University, Singapore. This research was partially funded by NTU's Research Scholarship and the National Research Foundation of Singapore under Grant NRF2011 NRFCRP001-090, Award Number NRF-CRP8-2011-03.

## References

[1] J. C. Ho, G. Jindal, M. Low, Singapore's intended nationally determined contribution for cop21 climate conference in paris, Energy Studies Institute, 2015.

[2] K. Li, C. Hu, G. Liu, W. Xue, Building's electricity consumption prediction using optimized artificial neural networks and principal component analysis, Energy and Buildings, ELSEVIER 108 (2015) 106-113.

[3] A. Kusiak, M. Li, Cooling output optimization of an air handling unit, Applied Energy, ELSEVIER (2010) 901-909.

[4] A. Arabali, M. Ghofrani, M. Etezadi-Amoli, M. S. Fadali, Y. Baghzouz, Genetic-algorithm-based optimization approach for energy management, IEEE Transactions on Power Delivery 28 (2012) 162-170.

[5] Y. Zhang, Indoor Air Quality Engineering, CRC Press, 2004.

[6] N. Klepeis, The national human activity pattern survey (nhaps) - a resource for assessing exposure to environmental pollutants, National Exposure Research Laboratory, U.S. Environmental Protection Agency.

[7] P. Fanger, Thermal comfort: Analysis and applications in environmental engineering, Danish Technical Press, 1970.

[8] ASHRAE, Standard 55-2013: Thermal environmental conditions for human occupancy.

[9] J. Xiong, X. Zhou, Z. Lian, J. You, Y. Lin, Thermal perception and skin temperature in different transient thermal environments in summer, Energy and Buildings, ELSEVIER 128 (2016) 155-163.

[10] W. Liu, Z. Lian, Q. Deng, Use of mean skin temperature in evaluation of individual thermal comfort for a person in a sleeping posture under steady thermal environment, Indoor and Built Environment, ELSEVIER 24 (2014) 489-499.

[11] S. Takada, S. Matsumoto, T. Matsushita, Prediction of whole-body thermal sensation in the non-steady state based on skin temperature, Building and Environment, ELSEVIER 68 (2013) 123-133.

[12] C. Dai, H. Zhang, E. Arens, Z. Lian, Machine learning approaches to predict thermal demands using skin temperatures: Steady-state conditions, Building and Environment, ELSEVIER 114 (2017) 1-10.

[13] D. Zhai, Y. C. Soh, W. Cai, Operating points as communication bridge between energy evaluation with air temperature and velocity based on extreme learning machine (elm) models, in: Proceedings of IEEE Conference on Industrial Electronics and Applications, 2016, pp. 712-716.

[14] D. Zhai, T. Chaudhuri, Y. C. Soh, Modeling and optimization of different sparse augmented firefly algorithms for acmv systems under two case studies, Building and Environment 125 (2017) 129-142.

[15] I. T. Michailidis, T. Schild, P. M. Roozbeh Sangi, C. Korkas, J. Ftterer, D. Mller, E. B. Kosmatopoulos, Energy-e_cient hvac management using cooperative, self-trained, control agents: A real-life german building case study, Applied Energy 211 (2018) 113-125.

[16] W. Wu, H. M. Skye, P. A. Domanski, Selecting hvac systems to achieve comfortable and cost-e_ective residential net-zero energy buildings, Applied Energy 212 (2018) 577-591.

[17] Y. He, M. Liu, T. Kvan, S. Peng, An enthalpy-based energy savings estimation method targeting thermal comfort level in naturally ventilated buildings in hot-humid summer zones, Applied Energy 187 (2017) 717-731.

[18] M. H. Fathollahzadeh, G. Heidarinejad, H. Pasdarshahri, Prediction of thermal comfort, iaq, and energy consumption in a dense occupancy environment with the under floor air distribution system, Building and Environment 90 (2015) 96-104.
[19] J.-H. Choi, D. Yeom, Study of data-driven thermal sensation prediction model as a function of local body skin temperatures in a built environment, Building and Environment, ELSEVIER 121 (2017) 130-147.
[20] Y. Wang, J. Kuckelkorn, F.-Y. Zhao, D. Liu, A. Kirschbaum, J.-L. Zhang, Evaluation on classroom thermal comfort and energy performance of passive school building by optimizing hvac control systems, Building and Environment, ELSEVIER 89 (2015) 86-106.
[21] K. S. Cetin, L. Manuel, A. Novoselac, Effect of technology-enabled time-of-use energy pricing on thermal comfort and energy use in mechanically conditioned residential buildings in cooling dominated climates, Building and Environment 96 (2016) 118-130.
[22] L. T. Wong, K. W. Mui, C. T. Cheung, Bayesian thermal comfort model, Building and Environment 82 (2014) 171-179.
[23] E. Brochu, V. M. Cora, N. de Freitas, A tutorial on bayesian optimization of expensive cost functions, with application to active user modeling and hierarchical reinforcement learning, arXiv.
[24] F. Ascione, N. Bianco, C. D. Stasioa, G. M. Mauro, G. P. Vanoli, Simulation-based model predictive control by the multi-objective optimization of building energy performance and thermal comfort, Energy and Buildings, ELSEVIER 111 (2016) 131-144.
[25] J.-M. Dussault, M. Sourbron, L. Gosselin, Reduced energy consumption and enhanced comfort with smart windows: Comparison between quasi-optimal, predictive and rule-based control strategies, Energy and Buildings 127 (2016) 680-691.
[26] X. Chen, Q. Wang, J. Srebric, Occupant feedback based model predictive control for thermal comfort and energy optimization: A chamber experimental evaluation, Applied Energy 164 (2016) 341-351.
[27] M. Pritoni, K. Salmon, A. Sanguinetti, J. Morejohn, M. Modera, Occupant thermal feedback for improved e_ciency in university buildings, Energy and Buildings 144 (2017) 241-250.
[28] J. Kim, T. Hong, J. Jeong, C. Koo, K. Jeong, An optimization model for selecting the optimal green systems by considering the thermal comfort and energy consumption, Applied Energy, ELSEVIER 169 (2016) 682-695.
[29] D. Zhai, Y. C. Soh, Balancing indoor thermal comfort and energy consumption of acmv systems via sparse swarm algorithms in optimizations, Energy and Buildings 149 (2017) 1-15.
[30] D. Zhai, Y. C. Soh, Balancing indoor thermal comfort and energy consumption of air-conditioning and mechanical ventilation systems via sparse firefly algorithm optimization, in: Proceedings of IEEE Joint Conference on Neural Networks, 2017, pp. 1488-1494.
[31] J. Babiak, B. W. Olesen, D. Petras, Low temperature heating and high temperature cooling, Guidebook No. 7, REHVA (Federation of European Heating and Air-conditioning Associations).
[32] C. E. Rasmussen, C. K. I. Williams, Gaussian Processes for Machine Learning., MIT Press, Cambridge, Massachusetts, 2006.

# Appendix

**Table 5: BGPO Evaluations for Case 1 and Case 2**

| Sample | ESR Mean (Case 1) | | ESR Mean (Case 2) | |
|---|---|---|---|---|
| | $\lambda \leq 0.3$ | $\lambda \geq 0.7$ | $\lambda \leq 0.3$ | $\lambda \geq 0.7$ |
| 10 | -0.0248 | **−0.2213** | 0.0382 | -0.1847 |
| 20 | -0.1429 | -0.1907 | 0.1135 | **−0.2112** |
| 30 | -0.1252 | -0.1827 | 0.1274 | **−0.2007** |
| 40 | -0.0820 | **−0.2043** | 0.0899 | -0.1783 |
| 50 | -0.1029 | **−0.2158** | -0.0475 | -0.2102 |

BGPO Mean Evaluations

| Sample | ESR Std Dev (Case 1) | | ESR Std Dev (Case 2) | |
|---|---|---|---|---|
| | $\lambda \leq 0.3$ | $\lambda \geq 0.7$ | $\lambda \leq 0.3$ | $\lambda \geq 0.7$ |
| 10 | 0.0237 | **0.0154** | 0.1549 | 0.0405 |
| 20 | 0.0487 | 0.0581 | 0.1784 | **0.0209** |
| 30 | 0.1141 | **0.0387** | 0.0833 | 0.0392 |
| 40 | 0.0700 | 0.0170 | 0.1253 | **0.0047** |
| 50 | 0.0874 | 0.0200 | 0.1229 | **0.0114** |

BGPO Standard Deviation Evaluations

**Table 6: AFA Evaluations for Case 1 and Case 2**

| Sample | ESR Mean (Case 1) | | ESR Mean (Case 2) | |
|---|---|---|---|---|
| | $\lambda \leq 0.3$ | $\lambda \geq 0.7$ | $\lambda \leq 0.3$ | $\lambda \geq 0.7$ |
| 10 | -0.0079 | -0.0850 | 0.0134 | **−0.0973** |
| 20 | -0.0286 | **−0.0998** | 0.0005 | -0.0926 |
| 30 | -0.0351 | -0.0949 | 0.0058 | **−0.0996** |
| 40 | -0.0199 | -0.0921 | 0.0449 | **−0.1007** |
| 50 | -0.0462 | -0.0920 | 0.0306 | **−0.1056** |

AFA Mean Evaluations

| Sample | ESR Std Dev (Case 1) | | ESR Std Dev (Case 2) | |
|---|---|---|---|---|
| | $\lambda \leq 0.3$ | $\lambda \geq 0.7$ | $\lambda \leq 0.3$ | $\lambda \geq 0.7$ |
| 10 | 0.0239 | 0.0119 | 0.0637 | **0.0056** |
| 20 | 0.0224 | **0.0061** | 0.0265 | 0.0210 |
| 30 | 0.0229 | 0.0175 | 0.0750 | **0.0060** |
| 40 | 0.0356 | 0.0270 | 0.0067 | **0.0056** |
| 50 | 0.0047 | 0.0213 | 0.0310 | **0.0000** |

AFA Standard Deviation Evaluations

**Table 7: BGPO and AFA Evaluations for Case 1 and Case 2 at Sample=50**

| $\lambda$ | | | 0 | 0.1 | 0.2 | 0.3 | 0.4 | 0.5 | 0.6 | 0.7 | 0.8 | 0.9 | 1 |
|---|---|---|---|---|---|---|---|---|---|---|---|---|---|
| Case 1 | BGPO | ESR | -0.1074 | -0.0265 | -0.0537 | -0.2239 | -0.1496 | -0.1947 | -0.2007 | -0.2133 | -0.1914 | -0.2183 | **−0.2401** |
| | | PMV | 0.0087 | 0.0110 | **0.0027** | -0.0313 | 0.0438 | -0.1204 | -0.0282 | -0.0058 | -0.0305 | -0.1738 | -0.1984 |
| | AFA | ESR | -0.0394 | -0.0495 | -0.0495 | -0.0464 | -0.0495 | -0.0495 | -0.0519 | -0.0602 | -0.1011 | -0.1011 | **−0.1056** |
| | | PMV | 0.0008 | **0.0007** | 0.0007 | -0.0011 | 0.0007 | 0.0007 | -0.0098 | -0.0778 | -0.4442 | -0.4442 | -0.4792 |
| Case 2 | BGPO | ESR | 0.1170 | -0.0403 | -0.1750 | -0.0914 | -0.1489 | -0.1838 | -0.2154 | -0.2166 | **−0.2224** | -0.1974 | -0.2044 |
| | | PMV | **0.3463** | 0.3472 | 0.3944 | 0.3758 | 0.4437 | 0.4469 | 0.5103 | 0.4715 | 0.4550 | 0.4595 | 0.4417 |
| | AFA | ESR | 0.0483 | 0.0416 | 0.0483 | -0.0156 | -0.1056 | -0.1056 | -0.1056 | -0.1056 | -0.1056 | -0.1056 | **−0.1056** |
| | | PMV | **0.1524** | 0.1640 | 0.1524 | 0.2570 | 0.3813 | 0.3813 | 0.3813 | 0.3813 | 0.3813 | 0.3813 | 0.3813 |